**Universal Design Principles for High-Quality Persistent Spin Textures**

*Cheng-Ao Ji, Lingling Tao, James M. Rondinelli, and Xue-Zeng Lu**

C.-A. Ji, X.-Z. Lu
Key Laboratory of Quantum Materials and Devices of Ministry of Education, School of Physics, Southeast University, Nanjing 211189, China
E-mail: xuezenglu@seu.edu.cn

L. L. Tao
School of Physics, Harbin Institute of Technology, Harbin 150001, China

J. M. Rondinelli
Department of Materials Science and Engineering, Northwestern University, Evanston, Illinois 60208, USA

Persistent spin texture (PST) describes a unique spin-momentum locking in momentum space that maintains a uniform spin orientation through portions of the Brillouin zone (BZ), enabling exceptionally long spin lifetimes which are essential for applications in spintronics. However, materials exhibiting large BZ regions of high-quality PST, characterized by minimal spin deviation and long spin lifetimes, remain scarce. Here a universal model is introduced to capture the formation of superior PST regions arising from the interplay of spin-orbit fields at different *k* points. Within this framework, high-quality PSTs are identified in several systems belonging to various point groups. Notably, the nonpolar-chiral compound $Na_2Sn_2O_3$ exhibits ~0.02 Å$^{-2}$ high-quality PST region, which can be reversed by the switching of geometric chirality, while $AgClO_4$ ($D_{2d}$ symmetry) exhibits a 0.016 Å$^{-2}$ PST region. Significantly, $Na_2Sn_2O_3$ and $AgClO_4$ host persistent spin helices with spin lifetimes of 0.5–7.4 ns and 0.9–2.5 ns, respectively, among the longest reported for PST materials. In addition, both chemical substitutions and the application of pressure are demonstrated as effective routes for engineering high-quality PST. Our findings not only establish a universal principle for high-quality PST, but also provide promising materials across various point groups for the next-generation spintronic devices.

## 1. Introduction

A spin-wave mode with an infinite spin lifetime, known as a persistent spin helix (PSH), arises from a persistent spin texture (PST) in momentum space through unidirectional spin-momentum locking.[1–5] The spin component of the PSH is protected by SU(2) symmetry, which is robust to spin-independent disorder.[1,6] This symmetry-driven suppression of spin relaxation makes PST an essential spin texture for enabling long coherence times for spintronic applications, including spin field effect transistors,[7] spin Hall effect,[8] and ferroelectric spin-orbit valve effect.[9]

Following the theoretical prediction in 2006, the PST was experimentally observed in quantum well structures of GaAs/AlGaAs and GaAs/InAlAs,[10–13] which originates from the fine balance between the strength of two spin-orbit coupling (SOC) effects, i.e., the Rashba[14] and Dresselhaus[15] interactions. Beyond this SOC-balancing mechanism, another distinct class of PSTs enforced by crystal symmetry has been identified in several polar materials, including $BiInO_3$, $A_3B_2O_7$ Ruddlesden-Popper family and 2D hybrid perovskites.[16–18] Furthermore, pure third-order SOC interactions, exclusively permitted in specific point groups, can also produce PSTs in both 3D and 2D systems, which possibly span the whole Brillouin zone (BZ).[19,20] Additionally, recent studies have found that in certain compounds with strong Rashba anisotropy, a PST can arise even along *k* paths lacking the mirror symmetry protection, a phenomenon referred to as accidental PST.[21,22]

Nevertheless, significant challenges remain in achieving high-quality PSTs and realizing its practical applications. First, a high-quality PST, according to the assessment criteria proposed in the previous study,[17] requires small spin deviation (less than 5°) and large PST area, which is in favor of large area of doping and long spin lifetime. While most compounds suffer from limited PST area and the substantial spin deviation, leading to scarcity of the materials exhibiting high-performing PST. Second, in quantum well structures, precise control over the well width and carrier density poses a major experimental hurdle. Furthermore, in many candidate materials, the PST region is located far from the valence or conduction band edges, complicating experimental realization and manipulation of PSTs. Consequently, there is great demand to identify and design high-quality PST compounds for next-generation spintronic devices.

Previous studies on PST have primarily focused on momentum-dependent spin-orbit fields around a single high-symmetry point.[18,19,23,24] This raises the question: can a PST emerge from the interplay of the spin-orbit fields associated with multiple distinct *k* points, and how would such PST differ from those governed by the single spin-orbit field? In this Letter, we address this gap by constructing universal models with $k{\cdot}p$ effective Hamiltonians that capture PST formation through interactions among distinct spin-orbit fields. Our framework identifies four classes of high-quality PSTs and predicts their occurrence in a wide range of materials belonging to various noncentrosymmetric point groups, including polar, nonpolar, chiral and achiral. Furthermore, we computationally show that chemical substitution and pressure provide effective strategies for tuning spin textures in part of compounds, thereby expanding the palette of materials exhibiting high-quality PSTs. Our study establishes a general design principle for discovery of high-quality PSTs, and proposes several candidates for integration into spintronic devices.

**2. PST formed by interactions of different spin-orbit fields.**

In systems without inversion symmetry, the SOC interaction produces a momentum-dependent effective magnetic field that interacts with the spin, resulting in momentum-dependent spin textures in reciprocal space. Three classes of spin textures, Rashba, Dresselhaus, and Weyl, are shown in **Figure 1**a-c. These correspond to three typical spin-orbit fields labelled in the figures by R, D, and W, respectively, emanating from different points in the BZ, and have been both theoretically predicted and experimentally observed in the materials belonging to diverse crystallographic point groups.[25–31] It is usually recognized that $\alpha = \beta$ is required for the typical spin-orbit field, where $\alpha$ and $\beta$ denote the SOC coefficients for the $k_x$ and $k_y$ terms, respectively. For instance, the Rashba spin-orbit field, $\boldsymbol{B_R} = \alpha_R(-k_y, k_x)$, requires $\alpha = \beta = \alpha_R$ for a spin-orbit field $\boldsymbol{B} = (-\beta k_y, \alpha k_x)$ (Figure 1a). When $\alpha \neq \beta$, the spin-orbit field can be described as the coexistence of two typical spin-orbit fields, e.g., $\boldsymbol{B} = (\alpha k_x, \beta k_y) = ((\alpha_W - \alpha_D)k_x, (\alpha_W + \alpha_D)k_y) = \boldsymbol{B_W} + \boldsymbol{B_D}$, which we refer to as the inhomogeneous spin-orbit field. Starting from a $k{\cdot}p$ effective Hamiltonian retaining only the linear terms, the interaction of two Weyl-type spin-orbit fields can be expressed as

$$H = (\boldsymbol{B}_1 + \boldsymbol{B_2}) \cdot \boldsymbol{\sigma} = (\alpha_1 - \alpha_2)k_x\sigma_x + (\beta_1 k_y + \beta_2 k_y')\sigma_y$$

defining a Weyl-Weyl (WW) model. Here $\boldsymbol{B_1}$ and $\boldsymbol{B_2}$ are momentum-dependent Weyl fields centered at distinct wavevectors $\boldsymbol{k_1}$ and $\boldsymbol{k_2}$, and all the components $k_x, k_y$ and $k_x, k_y'$ are positive, corresponding to the wavevectors associated with the $\boldsymbol{B_1}$ and $\boldsymbol{B_2}$ fields, respectively. $\alpha_i, \beta_i$ $(i = 1{,}2)$ denote the SOC coefficients for each field. Analogous, $k{\cdot}p$ effective Hamiltonians describing interactions between other types of spin-orbit fields are summarized in **Table 1**. Here the Rashba-Weyl model is excluded because it is not allowed along all $k$ paths of certain space groups (see Table S5, Supporting Information). The corresponding spin textures for these Hamiltonians are schematically shown in Figure 1d-i, labelled WW, DD-1, WD, RR, DD-2, RD, where Rashba, Dresselhaus or Weyl spin-orbit field contributes to the character for each wavevector in the pair of $k$ points.

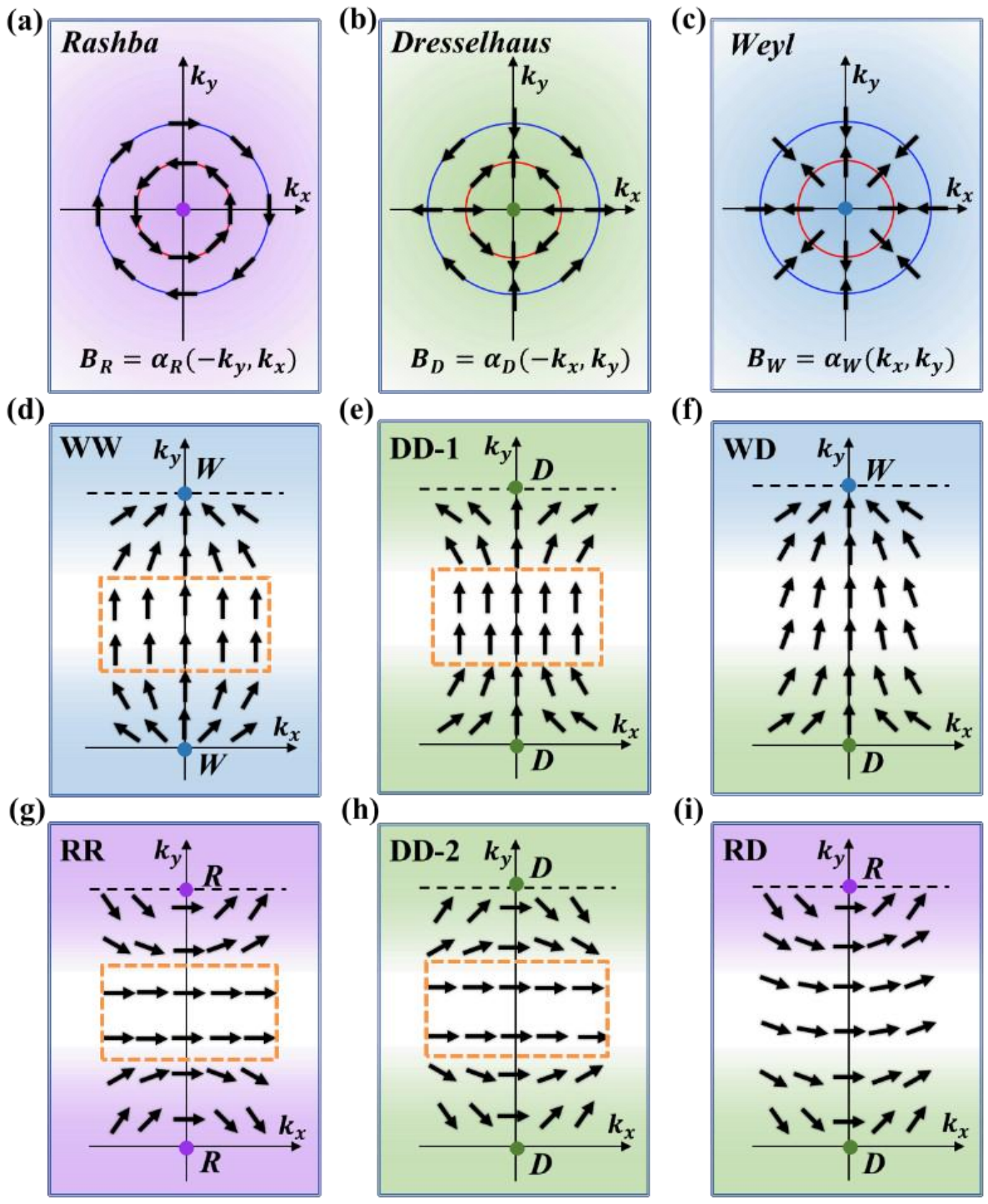

**Figure 1.** Schematic illustrations of spin textures in two-dimensional slices of the BZ. The (a)-(c) three typical classes of spin textures: Rashba-type (R), Dresselhaus-type (D) and Weyl-type (W), represented by purple, green and blue shading, respectively. Expressions for the three corresponding spin-orbit fields $\boldsymbol{B(k)}$ are shown as well. The (d)-(i) six kinds of spin textures described by our models generated by pairs of the typical fields: (d) WW, (e) DD-1, (f) WD, (g) RR, (h) DD-2, and (i) RD spin texture. The regions of PSTs are enclosed with orange lines.

Near a given $k$ point, the spin texture is dominated by the local single spin-orbit field described by one of the typical classes (Figure 1d-i), implying that $\alpha_1 \gg \alpha_2$ or vice versa. In contrast, midway along the path and far from two $k$ points, $\alpha_1 \approx \alpha_2$ naturally occurs within a certain range of momenta as the spin-texture character undergoes a transition. Interestingly, for the WW and DD-1 models with $\alpha_1 = \alpha_2$, the Hamiltonian reduces to $H = (\beta_1 k_y + \beta_2 k'_y)\sigma_y$, yielding zero spin deviation even when $k_x \neq 0$ (Table 1). This creates extended regions of high-quality PST in the BZ enclosed by orange dashed lines. This suggests that superior PST can expand across a large area even though the spin splitting is relatively small (here means $\beta_i$ is small). Similarly, the high-quality PST area also exists in the models of RR and DD-2, since only the spin components perpendicular to the $k$ path are allowed as $\alpha_1 = \alpha_2$. While according to derived Hamiltonian of the other two kinds of models (i.e., WD and RD), zero spin deviation is not allowed when $k_x \neq 0$ whatever the relation between $\alpha_1$ and $\alpha_2$, leading to less possibility for the discovery of superior PST regions. These findings are confirmed by the simulation of different Hamiltonians (Figures S1a-f, Supporting Information), which further shows that superior PSTs persist even for the inhomogeneous spin-orbit fields as in the case of the four prototypical models (Figures S1g-i, Supporting Information). As the ratio $\alpha_1/\alpha_2$ decreases, the PST regions for the four models can still remain sizable, unless that the ratio $\alpha_1/\alpha_2$ decreases by approximately two orders of magnitude where the region of PST is mainly induced by a single spin-orbit field (Figures S7 and S8, Supporting Information).

Moreover, as the cubic-in-*k* terms are included in the Hamiltonians, the zero spin deviation is still allowed in the four models provided that two requirements are satisfied: $\alpha_1 = \alpha_2$ and ${\alpha_{11}k_x}^2 + \alpha_{12}{k_y}^2 = \alpha_{21}{k_x}^2 + \alpha_{22}{k'_y}^2$ (see Tables S1 and S2, Supporting Information). As shown in Figures S2-S5, our simulations of spin textures parameterized with these models identify large PST regions. More generally, the spin textures obtained from the models including higher-order SOC terms can still be comparable to those computed by using the models including only the linear-in-*k* terms in the real materials (see Figure 3, Figure S6 and Table S3). This is attributed to the limited range of wavevector $k$ (~0.1 Å$^{-1}$) in the high-quality PST region. A high doping level may introduce electrons into the BZ regions where PST is absent and may also significantly influence the electronic structure, thereby affecting the spin lifetime of PSH.

Therefore, given the limited range of wavevector $k$ (~0.1 Å$^{-1}$) in the high-quality PST region, the effects of higher-order SOC terms on the spin deviation and the high-quality PST area are negligible. Then, the strategy for searching for high-quality PST is to minimize the spin deviation within such a limited range of wavevector used for carrier doping. Consequently, the linear-in-$k$ approximation is sufficient to describe the high-quality PST in our models, and thus provides a universal principle for finding superior PST materials across diverse point groups.

**Table 1.** $k{\cdot}p$ effective Hamiltonians for our models and the corresponding $S_{\mathrm{dv}}/S_{\mathrm{PST}}$. $\alpha_i, \beta_i$ $(i = 1{,}2)$ denote the SOC coefficients for different spin-orbit fields and the $\sigma_x$, $\sigma_y$ and $\sigma_z$ represent the Pauli matrices. $S_{\mathrm{dv}}$ and $S_{\mathrm{PST}}$ represent the expectation values of spin dominating spin deviation and PST, respectively.

| Type | Model | $S_{\mathrm{dv}}/S_{\mathrm{PST}}$ |
|---|---|---|
| WW | $(\alpha_1 - \alpha_2)k_x\sigma_x + (\beta_1 k_y + \beta_2 k'_y)\sigma_y$ | $\dfrac{(\alpha_1 - \alpha_2)k_x}{\beta_1 k_y + \beta_2 k'_y}$ |
| DD-1 | $(\alpha_2 - \alpha_1)k_x\sigma_x + (\beta_1 k_y + \beta_2 k'_y)\sigma_y$ | $\dfrac{(\alpha_2 - \alpha_1)k_x}{\beta_1 k_y + \beta_2 k'_y}$ |
| RR | $(\beta_1 k_y + \beta_2 k'_y)\sigma_x + (\alpha_2 - \alpha_1)k_x\sigma_y$ | $\dfrac{(\alpha_2 - \alpha_1)k_x}{\beta_1 k_y + \beta_2 k'_y}$ |
| DD-2 | $(\beta_1 k_y + \beta_2 k'_y)\sigma_x + (\alpha_1 - \alpha_2)k_x\sigma_y$ | $\dfrac{(\alpha_1 - \alpha_2)k_x}{\beta_1 k_y + \beta_2 k'_y}$ |
| WD | $(\alpha_1 + \alpha_2)k_x\sigma_x + (\beta_1 k_y + \beta_2 k'_y)\sigma_y$ | $\dfrac{(\alpha_1 + \alpha_2)k_x}{\beta_1 k_y + \beta_2 k'_y}$ |

| RD | $(\beta_1 k_y + \beta_2 k_y')\sigma_x + (\alpha_1 + \alpha_2)k_x\sigma_y$ | $\frac{(\alpha_1 + \alpha_2)k_x}{\beta_1 k_y + \beta_2 k_y'}$ |
|---|---|---|

Using symmetry analysis combined with $k{\cdot}p$ perturbation theory, we summarize and classify all types of spin textures in the $k_x$-$k_y$ plane about high-symmetry $k$ points with various little groups (Table S4, Supporting Information). These spin textures occur across all 21 noncentrosymmetric crystal classes, among which 13 point groups support $k$ paths consistent with our models. Guided by this framework, we identify all $k$ paths that enable high-quality PSTs in 109 crystalline space groups, listed in Table S5 (Supporting Information), facilitating the discovery of materials possessing superior PSTs. Notably, symmetry analysis reveals that some symmetry-protected PSTs reported in the previous studies are consistent with the RD model proposed in our work,[21] which usually does not support a large area of high-quality PST. In the following, based on the four models enabling large high-quality PST regions, we identified several compounds exhibiting superior PST regions around the CBM/VBM. In addition, some $k$ paths enabling high-quality PSTs lack any symmetry operations, and the allowed high-quality PSTs belong to the accidental PSTs that cannot be explained or predicted by conventional symmetry-protection principles.[21,22] Overall, our proposed model is a generalized framework which is useful for finding the high-quality PSTs, encompassing both symmetry-protected PST and accidental PST.

**3. Material candidates**

Based on the proposed models, we performed high-throughput screening of the Materials Project database through filtering the crystal space groups and the $k$-paths where the conduction band minimum (CBM) or valence band maximum (VBM) is located.[32] Then we identified candidate compounds with diverse point groups that may host high-quality PSTs (Figures S12a-b and Table S6, Supporting Information). Given that polar compounds have been studied extensively, here we focus on nonpolar chiral compounds of the form $A_2B_2O_3$ (B = Sn, Pb) crystallizing in the $I2_13$ space group. Symmetry analysis confirms that the Γ and H points in the BZ of this family (Figure S9b and Table S7, Supporting Information) belong to little groups with $T$ and $O$ symmetry, respectively, both supporting Weyl-type spin textures.

As shown in **Figure 2**b-d, our density functional theory (DFT) calculations including SOC effect confirm the existence of large PST regions (orange arrows) in these $A_2B_2O_3$ compounds, consistent with the WW model. For further verification of our theory, we derive the spin texture of $Na_2Sn_2O_3$ based on the $k{\cdot}p$ Hamiltonian for the WW model (Figure 2a), which agrees well with the direct DFT calculations. The extent of the PST regions varies among the different compounds, indicating that chemical substitution is an effective route to tune PSTs. Importantly, for $Na_2Sn_2O_3$ (Figure 2b), the CBM is located within the PST region, suggesting the possibility of a high-performing PSH and efficient spin transport. Our DFT calculations also confirm the dynamical stability of $Na_2Sn_2O_3$ (Figure S9d, Supporting Information). Therefore, we focus on this compound in the following discussion to examine its electronic structure and PST characteristics in detail.

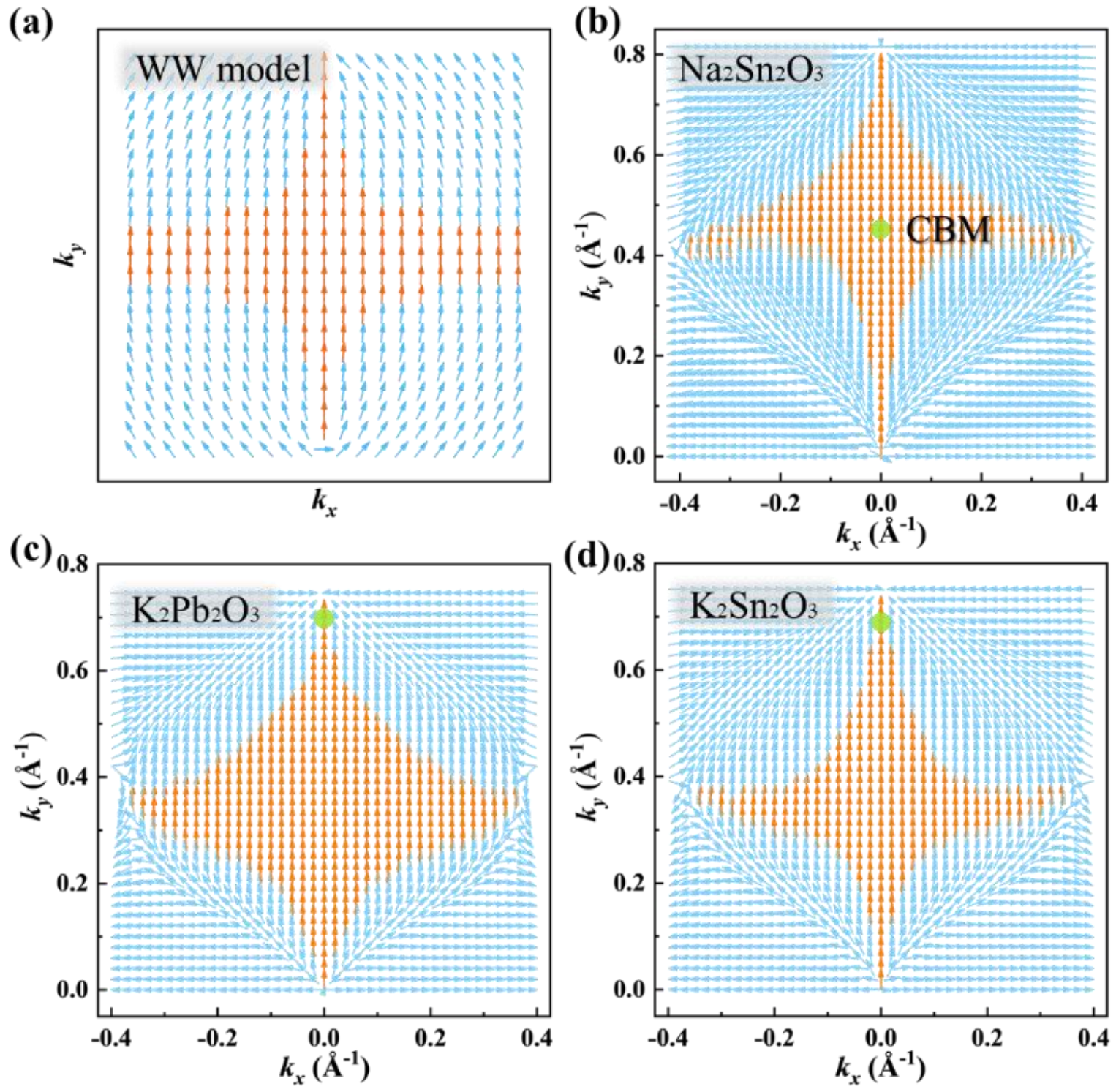

**Figure 2.** (a) Spin texture of $Na_2Sn_2O_3$ parameterized with the WW model. Here the parameters $\alpha_1, \beta_1$ and $\alpha_2, \beta_2$ are set as 0.43 eV Å and -1.28 eV Å, respectively, which are obtained by fitting band structures and spin textures around Γ and H (Table S8, Supporting Information). (b)-(d) The DFT+SOC spin textures for additional candidate compounds. PST regions with spin deviation less than 5° are indicated by orange arrows, while the remaining regions are marked with blue arrows. The location of the CBM is indicated by the green circle.

As presented in **Figure 3**a, the calculated band structure reveals that $Na_2Sn_2O_3$ is a semiconductor with a band gap of 80 meV. Importantly, it is found that the CBM for $Na_2Sn_2O_3$ is located midway along the Γ-H line, which is away from the spin-degenerate high-symmetry points; thus, satisfying the requirements for the realization of high-quality PST. Figure 3b shows that the area of the superior PST for electron doping is approximately 0.02 Å$^{-2}$ and corresponds to an energy of ~48 meV above the CBM. This finding demonstrates both the robustness of the PST and the validity of our theory.

In order to better describe the PST formed by interactions of different spin-orbit fields, we obtain SOC parameters for the $k{\cdot}p$ effective Hamiltonian capturing two portions of the PST region by fitting the band structure and spin configuration about the CBM (Table S8, Supporting Information). Corresponding to the portion of BZ where $k_y$ is larger than $k_{y-c}$ (the $k_y$ coordinate of the CBM), the dominant SOC coefficient, $\beta$ = 0.022 eV Å, which governs the PST direction, is nearly two orders of magnitude larger than $\alpha$ = 0.000192 eV Å, responsible for the spin deviation. This is similar for the BZ portion with $k_y$ smaller than $k_{y-c}$ ($\beta$ = -0.102 eV Å, $\alpha$ = 0.00156 eV Å). As can be seen in Figure 3c, the spin texture derived from this $k{\cdot}p$ Hamiltonian ($H = \alpha k_x \sigma_x + \beta k_y \sigma_y$) agrees well with the DFT results within the PST region. Solving the spin dynamics equation, we reveal that $Na_2Sn_2O_3$ exhibits an exceptionally long spin lifetime of approximately 0.5–7.4 ns for a Fermi wavelength of 0.067 $Å^{-1}$ and momentum-relaxation time of 1 ps, surpassing the spin lifetime of most previously reported PST compounds (Table S11, Supporting Information). The corresponding PSH yields a lifetime-to-period ratio of $1.51\times10^3$–$5.25\times10^3$ in $Na_2Sn_2O_3$ and this value is an order of magnitude larger than most typical PST materials (Table S10, Supporting Information). To the best of our knowledge, high-performing PSHs in nonpolar-chiral systems remain rare, and the discovery of such PSH in $Na_2Sn_2O_3$, guided by the proposed WW model, opens a promising avenue for their design.

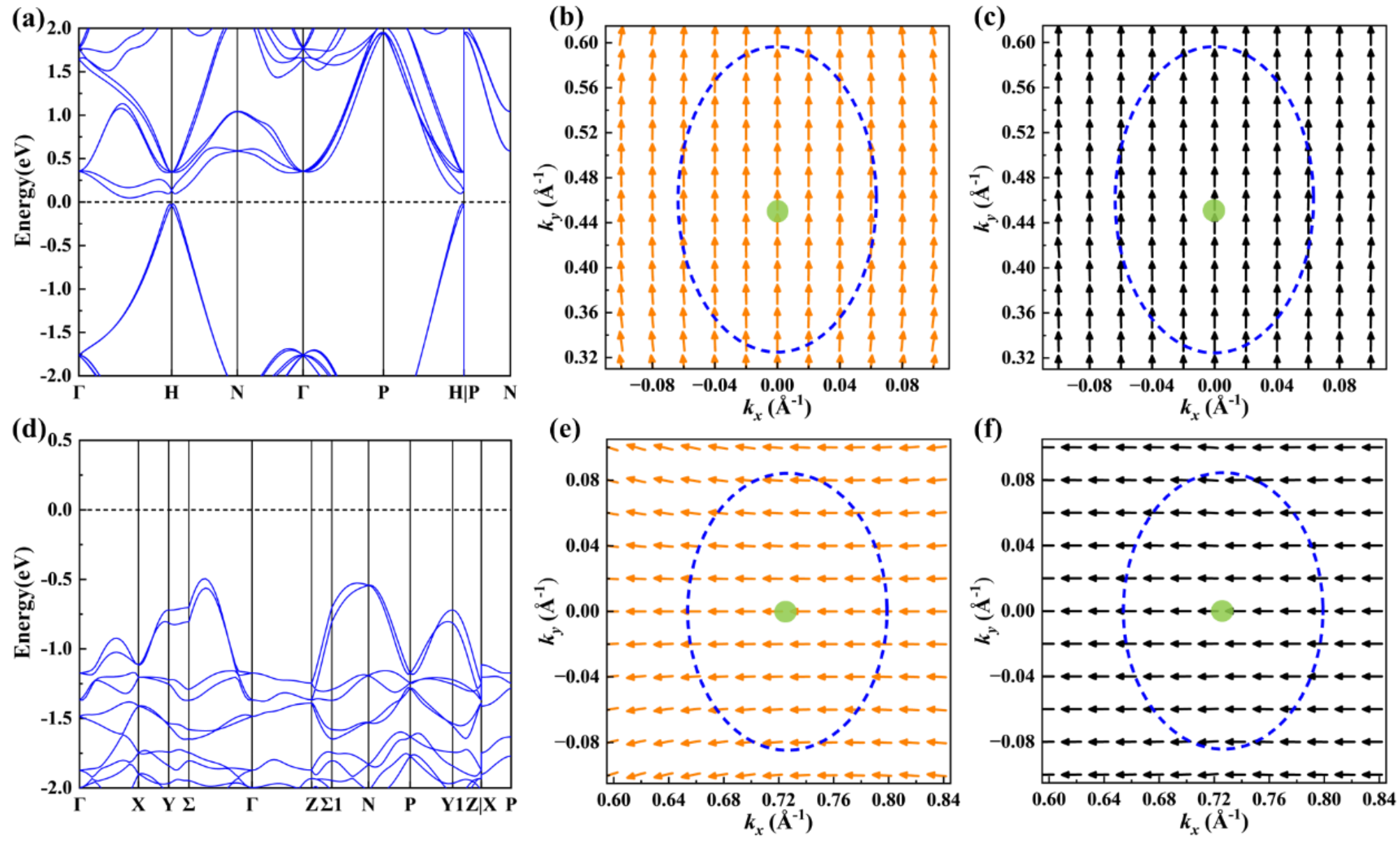

**Figure 3.** DFT+SOC band structures and spin textures in $k_x$-$k_y$ plane of (a, b) $Na_2Sn_2O_3$ and (d, e) $AgClO_4$. Spin textures based on the $k{\cdot}p$ models around (c) the CBM of $Na_2Sn_2O_3$ and (f) the VBM of $AgClO_4$. The high-quality PST regions for carrier doping are highlighted with the blue dashed circle. The location of CBM/VBM is indicated by the green circle.

It should be noted that the Fermi wavelength of 0.067 $Å^{-1}$ in $Na_2Sn_2O_3$ corresponds to a Fermi energy of 12.2 meV (electron-carrier density of $6\times10^{18}$ $cm^{-3}$), implying that the spin lifetime could be further enhanced at a lower doping level as was employed in the previous experiment (~$1.2\times10^{18}$ $cm^{-3}$).[33] In addition, reversing the structural chirality of $Na_2Sn_2O_3$ switches the spin-texture orientation (Figure S10, Supporting Information), enabling control over the direction of the spin texture via external fields such as electric impulse or electric fields that can toggle the geometric chirality.[34–36]

Aiming to demonstrate the generality of our theory, we examine a nonpolar-achiral yet noncentrosymmetric compound $AgClO_4$ with the $I\bar{4}2m$ space group. The silver perchlorate, with AgCl-subarray of the zincblende-type, has been synthesized at ambient conditions[37] and has semiconducting behavior with a band gap of about 2.3 eV (Figure 3d). The VBM is located midway along the Γ-Σ line and hole doping at ~32 meV below the VBM yields a high-quality PST region of nearly 0.016 $Å^{-2}$ (Figure 3e). Symmetry analysis indicates that the Γ point belongs to the $D_{2d}$ point group, producing a Dresselhaus-type spin texture, while the Σ point has $C_2$ symmetry, leading to an inhomogeneous spin texture.[25] From the SOC parameters obtained from the effective $k{\cdot}p$ Hamiltonian ($H = \alpha k_x\sigma_x + \beta k_y\sigma_y$) near the VBM (Table S9, Supporting Information), we further find that the Dresselhaus-type field dominates in the inhomogeneous spin-orbit field around Σ, leading to the high-quality PST consistent with the DD model (see Figures S1h and S11e, Supporting Information) and validating the generality of our theory. By fitting the band structure and spin configuration around the VBM, we obtain the SOC parameter $\alpha$ leading to PST, which is 0.102 eV Å and -0.0513 eV Å for the BZ portion of $k_x > k_{x-v}$ (the $k_x$ coordinate of the VBM) and $k_x < k_{x-v}$, respectively, far surpassing the value of $\beta$ (0.000656 eV Å for $k_x > k_{x-v}$, -0.000768 eV Å for $k_x < k_{x-v}$). This large Rashba anisotropy accounts for the large PST area with minimal spin deviation (Figure 3f). Remarkably, assuming a Fermi wavelength of 0.067 $Å^{-1}$ and a momentum-relaxation time of 1 ps, the computed spin lifetime of $AgClO_4$ reaches 0.9–2.5 ns, giving a lifetime-to-period ratio of $1.58 \times 10^3$–$8.39 \times 10^3$. This ratio is among the highest reported for PST compounds (Tables S10 and S11, Supporting Information).[17]

In order to further assess the practicality of the PST regions, we also investigated the PST regions for doping based on 3° and 7° spin deviation thresholds. It is found that PST area for hole doping in $AgClO_4$ continuously increases with the spin deviation, reaching ~0.032 $Å^{-2}$ when the spin deviation threshold is set to 7° (see Figures S13d-f, Supporting Information). Nevertheless, it should be noted that with more doping carriers, the spin lifetime will be reduced due to the existence of larger spin deviation.[17] For $Na_2Sn_2O_3$, the PST region for electron doping remains the same (~0.02 $Å^{-2}$) where the spin deviation is always less than 3°, and this is attributed to the distribution of band energy over $k$ points where the electrons are introduced into the area outside the PST region at higher doping levels, which will lead to a reduction in spin lifetime (see Figures S13a-c, Supporting Information).

Although our theory predicts numerous compounds with large PST regions with minimal spin deviations, some materials lack high-quality PSTs that we define as having small spin deviation less than 5° and large PST areas. This absence typically arises when the PST regions lie far from the CBM/VBM or when small spin splitting induces band crossings reversing the spin direction. In $AgClO_4$, applying hydrostatic pressure enhances the spin splitting and shifts the VBM closer to the center of the high-quality PST region (see Figure S14, Supporting

Information). As a result, the high-quality PST area for hole doping nearly doubles, and the spin lifetime extends to 4.0–30.8 ns at 3 GPa, giving rise to a large lifetime-to-period ratio of 1.17 $\times10^4$–2.02 $\times10^4$ (Tables S10 and S11, Supporting Information).

## 4. Conclusion

We have established a universal model describing interactions between various spin-orbit fields, enabling the realization of high-quality PSTs across a broad range of point groups. Guided by these design principles, we further discovered dynamically stable compounds exhibiting high-quality PST regions, consistent with our theory. Notably, the nonpolar-noncentrosymmetric compounds $Na_2Sn_2O_3$ and $AgClO_4$ exhibit PST areas of 0.02 Å$^{-2}$ and 0.016 Å$^{-2}$, respectively. In the nonpolar-chiral material $Na_2Sn_2O_3$, PST reversal can be achieved by switching geometric chirality, offering a route for control over spin-momentum locking. Both compounds exhibit exceptionally long PSH lifetimes (0.5–7.4 ns for $Na_2Sn_2O_3$ and 0.9–2.5 ns for $AgClO_4$), surpassing most previously reported PST systems. Furthermore, pressure and chemical substitution emerge as effective strategies to engineer spin textures in materials lacking intrinsic high-quality and functional PSTs, broadening the candidate pool.

To highlight the advantages of our approach, Table S12 (Supporting Information) summarizes the formation mechanisms and performance advantages of all different PST types. In contrast with the PSTs created from quantum wells structures, which require balancing Rashba and Dresselhaus SOC strengths, our model achieves PSTs without precise fine-tuning of SOC strength. Compared to symmetry-protected PST governed by a single spin-orbit field, the PST predicted here offers superior quality, characterized by minimal spin deviation, large spatial extent, and long spin lifetime, because zero spin deviation is allowed independent of momentum. Importantly, our design principles enable efficient discovery of high-quality PST in materials spanning diverse point groups.

Interestingly, our DFT calculations reveal that the PSTs in $Na_2Sn_2O_3$ and $AgClO_4$ exhibit 3D characteristics in momentum space over finite energy windows near the Fermi level (Figures S9c and S11c, Supporting Information), which is distinct from previously proposed 3D-PSTs around high-symmetry points,[38] offering more routes for 3D-PST exploration. Recent symmetry-based analysis further suggest that all noncentrosymmetric space groups except the trivial space group $P1$, permit PSTs in nonmagnetic solids.[39] Nevertheless, regardless of the extrinsic conditions such as carrier density and momentum-relaxation time, only the existence of PSTs cannot guarantee the high-performing PSH, since high Rashba anisotropy—the focus of our work—plays a key role in supporting the long spin lifetime of PSH.[22,33] Critically, our research provides the universal design rules, paving the way for discovering materials with superior PST for next-generation spintronic applications.

## 5. Experimental Section

Our calculations were based on density functional theory (DFT) implemented in the Vienna ab initio Simulation Package (VASP) code[40–42] within the generalized gradient approximation (GGA) using the revised Perdew-Burke-Ernzerhof functional for solids (PBE-sol).[43] We used a 550-eVplane-wave cutoff energy for all calculations and projector augmented wave method[44] with Na 2*p* and 3*s* electrons, Sn 4*d*, 5*s* and 5*p* electrons, Ag 4*d* and 5*s* electrons, Cl 3*s* and 3*p* electrons, K 3*s*, 4*s* and 3*p* electrons, Pb 5*d*, 6*s* and 6*p* electrons, V 3*p*, 3*d* and 4*s* electrons, Rh 4*p*, 4*d* and 5*s* electrons, P 3*s* and 3*p* electrons, Ca 3*s*, 4*s* and 3*p* electrons, Zr 4*s*, 5*s*, 4*p* and 4*d* electrons, Rb 4*s*, 5*s* and 4*p* electrons, Nb 4*p*, 5*s* and 4*d* electrons, Bi 6*s* and 6*p* electrons, O 2*s* and 2*p* electrons treated as valence states. Gaussian smearing (0.10 eV width) is used for the Brillouin-zone integrations. The *k*-point sampling was tested and converged for the different cells. The convergence thresholds for the electronic relaxation and structure relaxation are $10^{-6}$ eV and 10 meV/Å, respectively. By symmetry analysis and $k{\cdot}p$ perturbation theory, the $k{\cdot}p$ models about different high-symmetry points and lines are constructed (see Section VIII and Tables S13-24, Supporting Information).

**Supporting Information**

Supporting Information is available from the Wiley Online Library or from the author.

**Acknowledgements**

C.-A.J. and X.-Z.L. were supported by the National Natural Science Foundation of China (NSFC) under Grant No. 12474081, the open research fund of Key Laboratory of Quantum Materials and Devices (Southeast University), Ministry of Education, the Start-up Research Fund of Southeast University. J.M.R. was supported by the National Science Foundation (NSF) under DMR-2413680. DFT calculations were performed on high-performance computers, supported by the Big Data Computing Center of Southeast University.

# Supporting Information

**Universal Design Principles for High-Quality Persistent Spin Textures**

*Cheng-Ao Ji, Lingling Tao, James M. Rondinelli, and Xue-Zeng Lu**

C.-A. Ji, X.-Z. Lu
Key Laboratory of Quantum Materials and Devices of Ministry of Education, School of Physics, Southeast University, Nanjing 211189, China
E-mail: xuezenglu@seu.edu.cn

L. L. Tao
School of Physics, Harbin Institute of Technology, Harbin 150001, China

J. M. Rondinelli
Department of Materials Science and Engineering, Northwestern University, Evanston, Illinois 60208, USA

## I. Our models

Considering the coupling of two kinds of momentum-dependent spin-orbit fields around different $k$ points, we obtain a series of $k \cdot p$ effective Hamiltonians, which may occur in various crystallographic point groups. Note that in the framework of our models, only the linear terms inducing the in-plane spin texture are included in the Hamiltonians. In the present work, six possible models are studied specifically. These models are adaptive to the spin patterns in the $k_x$-$k_y$ plane of the systems with diverse point groups, which is then demonstrated in our calculations.

As shown in Figure S1, we obtain a series of spin textures corresponding to different models through parameter simulation, where the parameters belonging to the same spin-orbit field is assumed to be equal. For better demonstration of our models, some other situations when the inequality of parameters exists (i.e., $\alpha_i \neq \beta_i$) are explored as well, which correspond to the $WW_{in}$, $DD_{in}$ and $RR_{in}$ models shown in Figures S1g-i (here the index 'in' means inhomogeneous). The $WW_{in}$ model means that one of the spin-orbit fields is the typical Weyl-type field while the other is an inhomogeneous spin-orbit field where the Weyl-type field take the lead, and similarly for the $RR_{in}$ and $DD_{in}$ models, Rashba-type and Dresselhaus-type field dominates in the inhomogeneous spin-orbit field, respectively. It is obvious that the high-quality PSTs are still allowed in the case of inhomogeneous spin-orbit fields.

### The effect of higher-order SOC terms

As shown in Table S2, according to the diverse SOC terms permitted by different point group symmetry (Table S1), we summarized the $k \cdot p$ effective Hamiltonians including linear and cubic terms for four models allowing the appearance of the high-quality PST (i.e., WW, RR, DD-1 and DD-2 models). We find that the zero spin deviation is still allowed as long as two requirements are satisfied, i.e., $\alpha_1 = \alpha_2$ and $\alpha_{11}{k_x}^2 + \alpha_{12}{k_y}^2 = \alpha_{21}{k_x}^2 + \alpha_{22}{k'_y}^2$. Essentially, as depicted in Figure S2, our simulation of spin textures based on the Hamiltonians reveals that large PST regions remain as cubic terms are included in the models. In addition, for the WW and RR models, with the change of the relationship between the SOC parameters for linear terms and those for cubic terms, the PST regions almost keep the same (see Figure S3 and Figure S4). Similar conclusions also apply to DD-1 and DD-2 models since the Hamiltonians of these models share the similar origin. For some other cases when $\alpha_1 \neq \beta_1$ or $\alpha_{11} \neq \alpha_{12}$, as displayed in Figure S5, the PST still persists in a large area.

For some specific point groups (i.e., $C_{3v}$ and $D_3$), the cubic terms will introduce the out-of-plane spin components in the $k_x$-$k_y$ plane (see Table S1). For some $k$ planes protected by the $C_2T$ symmetry, only the in-plane spin components can exist, which is consistent with our models. While for other $k$ planes without the perpendicular $C_2$ rotation axis or those with the perpendicular screw axis (i.e., the $C_2$ rotation axis combined with translation operation), out-of-plane spin components will be allowed. A high-quality PST may also exist in this case as the SOC coefficients leading to the out-of-plane spin components are relatively small compared with those dominating the in-plane spin components.[2]

Furthermore, by fitting the band structures and spin textures, we obtain a series of SOC parameters for linear and cubic terms in $Na_2Sn_2O_3$ and $AgClO_4$ (Table S3). Based on these SOC parameters, we simulated the high-quality PST regions around the CBM and VBM. As illustrated in Figure S6, the spin textures obtained from the models including higher-order SOC terms are comparable to those computed by using the models including only the linear-in-$k$ terms. This is attributed to the limited range of wavevector $k$ (~0.1 Å$^{-1}$) in the high-quality PST region, which corresponds to a carrier density of ~$10^{18}$ cm$^{-3}$–$10^{19}$ cm$^{-3}$. A high doping level may introduce electrons into the Brillouin zone (BZ) regions where PST is absent and may also significantly influence the electronic structure, thereby affecting the spin lifetime of PSH.

In the practical application of PST, given the limited range of wavevector $k$ (~0.1 Å$^{-1}$) in the high-quality PST region, the effects of higher-order SOC terms on the spin deviation and the high-quality PST area are negligible. Then, the strategy for searching for high-quality PST is to minimize the spin deviation within such a limited range of wavevector used for carrier doping. Consequently, the linear-in-$k$ approximation in the $k{\cdot}p$ model is sufficient to describe the high-quality PST in our models, and thus provides a universal principle for finding superior PST materials across diverse point groups.

## II. Quantitative analysis of the sensitivity of spin deviation to SOC parameters

**WW and DD-2 models**

As shown in Table 1, for the WW and DD-2 models, we have

$$S_{dv}/S_{PST} = \frac{(\alpha_1-\alpha_2)k_x}{\beta_1 k_y+\beta_2 k'_y} = \tan\delta\theta,$$

here the δθ is the spin deviation away from the PST direction. Then we can derive

$$\delta\theta = \arctan\left[\frac{(\alpha_1-\alpha_2)k_x}{\beta_1 k_y+\beta_2 k'_y}\right],$$

In order to realize the spin deviation less than 5°, the SOC coefficients $\lambda_1$ and $\lambda_2$ (here we assume $\alpha_1 = \beta_1 = \lambda_1$, $\alpha_2 = \beta_2 = \lambda_2$) should satisfy the following equations:

1) As $\frac{k_x}{k_y} \leq -\tan\ 5°$ and $\frac{k_x}{k'_y} \leq -\tan\ 5°$, $\frac{k_x+k'_y*\tan 5°}{k_x-k_y*\tan 5°} \leq \frac{\lambda_1}{\lambda_2} \leq \frac{k_x-k'_y*\tan 5°}{k_x+k_y*\tan 5°}$.

2) As $\frac{k_x}{k_y} \leq -\tan\ 5°$ and $\frac{k_x}{k'_y} \geq -\tan\ 5°$, $\frac{\lambda_1}{\lambda_2} \leq \frac{k_x-k'_y*\tan 5°}{k_x+k_y*\tan 5°}$.

3) As $\frac{k_x}{k_y} \geq -\tan\ 5°$ and $\frac{k_x}{k'_y} \leq -\tan\ 5°$, $\frac{\lambda_1}{\lambda_2} \geq \frac{k_x+k'_y*\tan 5°}{k_x-k_y*\tan 5°}$.

4) As $-\tan\ 5° \leq \frac{k_x}{k_y} \leq \tan\ 5°$ and $-\tan\ 5° \leq \frac{k_x}{k'_y} \leq \tan\ 5°$, the spin deviation δθ is always less than 5°.

5) As $\frac{k_x}{k_y} \geq \tan\ 5°$ and $\frac{k_x}{k'_y} \leq \tan\ 5°$, $\frac{\lambda_1}{\lambda_2} \leq \frac{k_x+k'_y*\tan 5°}{k_x-k_y*\tan 5°}$.

6) As $\frac{k_x}{k_y} \leq \tan\ 5°$ and $\frac{k_x}{k'_y} \geq \tan\ 5°$, $\frac{\lambda_1}{\lambda_2} \geq \frac{k_x-k'_y*\tan 5°}{k_x+k_y*\tan 5°}$.

7) As $\frac{k_x}{k_y} \geq \tan\ 5°$ and $\frac{k_x}{k'_y} \geq \tan\ 5°$, $\frac{k_x-k'_y*\tan 5°}{k_x+k_y*\tan 5°} \leq \frac{\lambda_1}{\lambda_2} \leq \frac{k_x+k'_y*\tan 5°}{k_x-k_y*\tan 5°}$.

Here $\lambda_1 = \frac{C_1}{\sqrt{{k_x}^2+{k_y}^2}}$ and $\lambda_2 = \frac{C_2}{\sqrt{{k_x}^2+{k'_y}^2}}$ ($C_1$ and $C_2$ are constants), which guarantees the normalization of the total spin expectation value induced by one spin-orbit field (i.e., ${S_{\mathrm{dv}}}^2 + {S_{\mathrm{PST}}}^2 = constant$ for the separate spin-orbit field). The equations above give the range of $\lambda_1/\lambda_2$ (i.e., $\alpha_1/\alpha_2$) for different $k_x$ and $k_y$, which corresponds to the seven districts in the momentum space shown in Figure S7. In order to quantify the effect of the $\alpha_1/\alpha_2$ ratio on the spin deviation and PST area, based on these seven equations, we simulated the PST regions with spin deviation less than 5° by setting different values of $C_1/C_2$ (Figure S8). As shown in Figures S8a-b, when $C_1$ is comparable to $C_2$, there always exists a large area of PST. As $C_2$ continues to increase, approaching a value at most two orders of magnitude larger than $C_1$, the PST region still remains sizable and becomes closer to one of the high-symmetry $k$ points, which is mainly induced by a single spin-orbit field (see Figures S8c-d).

**RR and DD-1 models**

Similarly, for the RR and DD-1 models, we have

$$S_{dv}/S_{PST} = \frac{(\alpha_2-\alpha_1)k_x}{\beta_1 k_y+\beta_2 k_y'} = \tan\delta\theta,$$

the SOC coefficients $\lambda_1$ and $\lambda_2$ (here we assume $\alpha_1 = \beta_1 = \lambda_1$, $\alpha_2 = \beta_2 = \lambda_2$) should satisfy the following equations:

1) As $\frac{k_x}{k_y'} \leq -\tan\ 5°$ and $\frac{k_x}{k_y} \leq -\tan\ 5°$, $\frac{k_x+k_y*\tan 5°}{k_x-k_y'*\tan 5°} \leq \frac{\lambda_2}{\lambda_1} \leq \frac{k_x-k_y*\tan 5°}{k_x+k_y'*\tan 5°}$.

2) As $\frac{k_x}{k_y'} \leq -\tan\ 5°$ and $\frac{k_x}{k_y} \geq -\tan\ 5°$, $\frac{\lambda_2}{\lambda_1} \leq \frac{k_x-k_y*\tan 5°}{k_x+k_y'*\tan 5°}$.

3) As $\frac{k_x}{k_y'} \geq -\tan\ 5°$ and $\frac{k_x}{k_y} \leq -\tan\ 5°$, $\frac{\lambda_2}{\lambda_1} \geq \frac{k_x+k_y*\tan 5°}{k_x-k_y'*\tan 5°}$.

4) As $-\tan\ 5° \leq \frac{k_x}{k_y'} \leq \tan\ 5°$ and $-\tan\ 5° \leq \frac{k_x}{k_y} \leq \tan\ 5°$, the spin deviation $\delta\theta$ is always less than 5°.

5) As $\frac{k_x}{k_y'} \geq \tan\ 5°$ and $\frac{k_x}{k_y} \leq \tan\ 5°$, $\frac{\lambda_2}{\lambda_1} \leq \frac{k_x+k_y*\tan 5°}{k_x-k_y'*\tan 5°}$.

6) As $\frac{k_x}{k_y'} \leq \tan\ 5°$ and $\frac{k_x}{k_y} \geq \tan\ 5°$, $\frac{\lambda_2}{\lambda_1} \geq \frac{k_x-k_y*\tan 5°}{k_x+k_y'*\tan 5°}$.

7) As $\frac{k_x}{k_y'} \geq \tan\ 5°$ and $\frac{k_x}{k_y} \geq \tan\ 5°$, $\frac{k_x-k_y*\tan 5°}{k_x+k_y'*\tan 5°} \leq \frac{\lambda_2}{\lambda_1} \leq \frac{k_x+k_y*\tan 5°}{k_x-k_y'*\tan 5°}$.

These equations above have the similar formation with those of the WW and DD-2 models, thus the similar conclusion can be obtained. Essentially, in contrast to the RD and WD models, the PST in the WW, RR, DD-1 and DD-2 models always persists in a larger region irrespective of the changes in $C_1/C_2$, when the same SOC parameters are used.

## III. Materials screening

Based on our models, we screen the database and obtain a series of materials families, which may present high-quality PST. The screening procedure is as follows: (1) Combining the previous studies, we identify the wavevector point groups (little groups) which may present typical spin-orbit fields. (2) We then identify the space groups where these wavevector point groups exist. (3) Then in every space group, based on symmetry analysis, we summarize the $k$ paths in which the $k$ points on the both sides may present typical spin-orbit fields satisfying our models. (4) We screen the database using these constraints and identify compounds for which the CBM/VBM is located on the required $k$ paths from Step (3).

Note that in some of the $k$ paths, where only one of the $k$ points present typical spin-orbit field or both $k$ points present the inhomogeneous spin-orbit fields, the high-quality PST may exist according to our theory. For some $k$ planes where the CBM/VBM is located, only the in-plane spin components can exist due to the protection of $C_2T$ symmetry, and this situation agrees with our models. While for the $k$ planes lacking the perpendicular $C_2$ rotation axis or some special $k$ planes with the perpendicular screw axis (i.e., the $C_2$ rotation axis combined with translation operation), terms such as $\sigma_z k_x$ or $\sigma_z k_y$ will be allowed, leading to out-of-plane spin components.[1] A high-quality PST may exist in this case as well, which is usually attributed to the large Rashba anisotropy or the relatively small magnitude of the SOC coefficients dominating the out-of-plane spin components.

Only a subset of materials is listed in Table S6, where the CBM/VBM is close to the midway of the required $k$ paths summarized in Table S5. Many other compounds host high-quality PSTs even when the CBM/VBM is offset from these midpoints. Such cases can be engineered through external forces, e.g., applying strain or pressure,[2] to modify the band dispersion and SOC strength, thereby shifting the CBM/VBM toward the optimal PST region.

## IV. Crystal structures, band structures and spin textures

We primarily concentrate on the band structures of two kinds of nonpolar materials (i.e., $Na_2Sn_2O_3$ and $AgClO_4$). The crystal structures and the corresponding BZ of the two compounds are depicted in Figures S9a-b and Figures S11a-b, respectively, with the high symmetry points listed in Table S7.

Furthermore, we also searched for some other compounds in different noncentrosymmetric point groups (including polar or nonpolar, chiral or achiral) to validate our models. All the materials are dynamically stable and the characteristics of these compounds, including quality of the PSTs, space groups, and band gaps, are summarized in Table S6. As can be seen in Figure S12, NaSnP and $RbBiNb_2O_7$ both exhibit high-quality PSTs with large areas, which are in agreement with the RR and DD-2 models of our theory. In contrast, for VSnRh and $Ca_3Zr_2O_7$, the PST regions with minimal spin deviation are relatively limited, which correspond to the WD and RD models. Notably, the PST area of $RbBiNb_2O_7$ is further enhanced, compared with that studied in $CsBiNb_2O_7$, which is attributed to the change of VBM, indicating that chemical substitution is an effective method of tuning the PST.[3,4]

### V. Comparison of PST areas with different spin deviation threshold

In order to better assess the practicality of the PST regions, as shown in Figure S13, we computed the PST regions with different choice of spin deviation threshold. It is obvious that, for both $Na_2Sn_2O_3$ and $AgClO_4$, the areas of the PST regions (indicated by orange arrows) gradually increase with a larger spin deviation threshold. By further considering the distribution of band energy over *k* points, we confirmed the high-quality PST area (enclosed by blue circle) which can be used for carrier doping. It is discovered that PST area in $AgClO_4$ continuously increases with the spin deviation, reaching ~0.032 $Å^{-2}$ when the spin deviation threshold is set to 7° (Figures S13d-f). For $Na_2Sn_2O_3$, the PST region for electron doping remains the same (~0.02 $Å^{-2}$) where the spin deviation is always less than 3°, and this is attributed to the distribution of band energy over *k* points (Figures S13a-c). As shown in Figures S13b-c, at higher doping levels, electrons are introduced into the area outside the PST region (enclosed by black circle), which will lead to a reduction in spin lifetime.

With more doping carriers, the spin lifetime will be reduced due to the existence of larger spin deviation.[5] Besides, excessive doping may also have effects on electronic structures, thereby affecting the quality of PST. Considering a fine balance between spin lifetime and doping level, we choose 5° spin deviation threshold as the assessment criterion. For $Na_2Sn_2O_3$ and $AgClO_4$, the high-quality PST areas with spin deviation less than 5° reach 0.02 $Å^{-2}$ and 0.016 $Å^{-2}$, respectively, corresponding to the carrier densities of $6\times10^{18}$ $cm^{-3}$ and $2.6\times10^{19}$ $cm^{-3}$, which are comparable to those typically employed in experiments.[6–8]

## VI. Pressure regulation

For some of the materials lacking superior PSTs, pressure may be used to fine tune the spin textures. Here we mainly discuss pressure effects on the properties of $AgClO_4$, where the VBM is not located at the centre of the PST region, leading to the moderate but limited area of high-quality PST shown in Figure S14b. By imposing pressure of 5 GPa (Figure S14a), we enhance the spin splitting from 67.6 meV to 99.5 meV. Significantly, when the employed pressure is 3 GPa, which is the lower bound for the critical pressure for the phase transition,[9] the area of high-quality PST is nearly doubled and the spin lifetime is extended as well (Figure S14c).

## VII. Comparison of different PSTs

Aimed at emphasizing the superior quality of the PST mentioned in this work, we also compare the PST formed by the interactions of two effective spin-orbit fields with other types of PSTs studied earlier (Table S12). In contrast with the PST in the quantum wells caused by the balance of Rashba and Dresselhaus SOC effects, the existence of PST described in this work has no need for manipulation of the SOC strength since it is protected by the intrinsic symmetry and formed by interactions of two spin-orbit fields. Compared with the symmetry-protected PST governed by the single spin-orbit field, the proposed PST in this work usually exhibits high quality with minimal spin deviation, large area and long spin lifetime. On one hand, because part of the linear-in-$k$ terms is compensated by interactions between two spin-orbit fields, the spin deviation mainly originates from the cubic-in-$k$ terms which are relatively small. On the other hand, the PST proposed in this work tends to occur at the midpoint of high-symmetry lines, which is far from the spin-degenerate high symmetry points, usually leading to large PST area. Furthermore, according to our theory, a high-quality PST is allowed in systems with various kinds of point groups, which provides more routes for finding candidate materials.

## VIII. $\boldsymbol{k}\cdot\boldsymbol{p}$ model and spin diffusion equation

By symmetry analysis and *k·p* perturbation theory, the *k·p* models about different high-symmetry points and lines are constructed, and the parameters of different terms in the models are derived by fitting the band structure and spin texture obtained from DFT calculations, which are summarized in Tables S5 and S6.

### Derivation of $\boldsymbol{k}\cdot\boldsymbol{p}$ model

If including the spin ($|\boldsymbol{S}|$=1/2) in the wavefunction, there will be a spin-orbital coupling (SOC) term $\propto \boldsymbol{s}\cdot\boldsymbol{l}$ where $\boldsymbol{s}$ and $\boldsymbol{l}$ are spin and orbital angular momenta, respectively, in $\mathcal{H}$. $\mathcal{H}(\Delta\boldsymbol{k})$ becomes:[10,11]

$$\mathcal{H}(\Delta\boldsymbol{k}) = \mathcal{H} + \frac{\hbar\boldsymbol{\Delta k}}{m}\cdot\left(\boldsymbol{p} + \frac{\hbar}{4mc^2}\boldsymbol{\sigma}\times\boldsymbol{\nabla V}\right) + \frac{\hbar^2}{2m}(\Delta\boldsymbol{k})^2 \qquad \text{(S1)}$$

$\mathcal{H}(\Delta\boldsymbol{k})$ in (S1) can be solved using a group-theory analysis with the following steps. Here we solve (S1), as an example, and set $\boldsymbol{k}_0 = 0$:

(1) Choosing a basis set of wavefunctions, e.g., $D^{\frac{3}{2}}$ can be used in a four-band $\boldsymbol{k}\cdot\boldsymbol{p}$ model.

(2) Decomposing $D^{\frac{3}{2}}$ into a sum of irreducible representations (*irreps*) (e.g., $\mathcal{R}$) using the character table of a double group at $\boldsymbol{k}_0$.

After accomplishing (1) and (2), the irreducible symmetric matrix can be obtained by considering

(3) $\mathcal{H}_{i,j}^{\alpha,\beta}(\boldsymbol{k}) = \sum_{\gamma,l}(\psi_i^{\alpha}, \mathcal{H}_l^{\bar{\gamma}}\psi_j^{\beta}) = \sum_{\gamma,l}\left(\begin{matrix}\alpha\\ i\end{matrix}\tau\middle|\begin{matrix}\bar{\gamma}\beta\\ l\,j\end{matrix}\right)\left\|\mathcal{H}_l^{\bar{\gamma}}(\boldsymbol{k})\right\|$, where $\mathcal{H}_{i,j}^{\alpha,\beta}(\boldsymbol{k})$ comprises the matrix of the $\boldsymbol{k}\cdot\boldsymbol{p}$ model and $\boldsymbol{k}$ is a small wavevector away from $\boldsymbol{k}_0$, $\psi_i^{\alpha}$ and $\psi_j^{\beta}$ are $D^{\frac{3}{2}}$ wavefunctions, $\mathcal{H}_l^{\bar{\gamma}}$ is the Taylor expansion of $\boldsymbol{k}$ where $\bar{\gamma}$ and $l$ indicate the conjugate *irreps* of the summations over the *irreps* obtained by decomposing $\mathcal{R}\times\mathcal{R}$ and the dimension of $\mathcal{H}^{\bar{\gamma}}$, respectively. $\left(\begin{matrix}\alpha\\ i\end{matrix}\tau\middle|\begin{matrix}\bar{\gamma}\beta\\ l\,j\end{matrix}\right)$ are the Clebsch-Gordon (C-G) coefficients and

$\|\mathcal{H}_l^{\bar{\gamma}}(\boldsymbol{k})\|$ is the irreducible symmetric tensors of $\boldsymbol{k}$ that can be obtained by using the projection operator technique.

(4) Last, the final Hamiltonian $\mathcal{H} = \sum_{\gamma,l} X_l^{\bar{\gamma}} ||H_l^{\bar{\gamma}}(\boldsymbol{k})||$, where $X_l^{\bar{\gamma}} = \begin{pmatrix} \alpha \\ i \end{pmatrix.\tau \left| \begin{matrix} \bar{\gamma}\beta \\ l\,j \end{matrix} \right)$ is the irreducible symmetric matrix.

Here we first formulate the $\boldsymbol{k} \cdot \boldsymbol{p}$ model around the high-symmetry $\boldsymbol{k}$-points Γ and H with *T* symmetry in the system $Na_2Sn_2O_3$. Because there are only one-dimensional *irreps*, we choose a basis set of wavefunctions having a symmetry of $D^{\frac{1}{2}}$ (Tables S13-S14), in which $D^{\frac{1}{2}} = D^{\frac{1}{2}} \times A = \bar{E}$. According to Step (2), we can have:

$$\overline{\overline{D^{\frac{1}{2}} \times D^{\frac{1}{2}}}} = \overline{\overline{\bar{E} \times \bar{E}}} = A + T \qquad \text{(S2)}$$

Then according to the symmetries of $A$ and $T$, we need to derive the symmetry-adapted $k$ terms and C-G coefficients, which are presented in Tables S15 and S16, respectively.

Now, according to Step (4) the Hamiltonian at Γ and H with *T* symmetry, where we consider the spin-orbit coupling terms constrained in the $k_x - k_y$ plane, can be expressed as:

$$H = \beta\left(\sigma_x k_x + \sigma_y k_y\right) \qquad \text{(S3)}$$

Here, it should be noted that the C-G coefficients have no direction. The way to assign the $\sigma_i$ ($i = x, y, z$) coupled to $k$ components is according to the structural symmetry. One can also understand this transformation as a change between the two different representations. Another rule to build the $\boldsymbol{k} \cdot \boldsymbol{p}$ model is that the final Hamiltonian should also fulfill time reversal symmetry.

The $\boldsymbol{k} \cdot \boldsymbol{p}$ model around the high-symmetry $\boldsymbol{k}$-point with $D_{2d}$ symmetry (here, at Γ in $AgClO_4$) can be obtained in a similar way. Because there are only one-dimensional *irreps*, we choose a basis set of wavefunctions having a symmetry of $D^{\frac{1}{2}}$ (Tables S17-S18), in which $D^{\frac{1}{2}} = D^{\frac{1}{2}} \times A_1 = \overline{E}_1$. According to Step (2), we have:

$$\overline{\overline{D^{\frac{1}{2}} \times D^{\frac{1}{2}}}} = \overline{\overline{\overline{E}_1 \times \overline{E}_1}} = A_1 + A_2 + E \qquad \text{(S4)}$$

Then according to the symmetries of $A$ and $B$, we need to derive the symmetry-adapted $k$ terms and C-G coefficients, which are presented in Tables S19 and S20, respectively.

Now, according to Step (4) the Hamiltonian at Γ with $D_{2d}$ symmetry, where we consider the spin-orbit coupling terms constrained in the $k_x - k_y$ plane, can be expressed as:

$$H = \beta(\sigma_x k_x - \sigma_y k_y) \qquad \text{(S5)}$$

The $\boldsymbol{k} \cdot \boldsymbol{p}$ model around the high-symmetry $\boldsymbol{k}$-point with $C_{2y}$ symmetry (here, it is the CBM (0, $k_y$, 0) in $Na_2Sn_2O_3$) can be obtained in a similar way. Because there are only one-dimensional *irreps*, we choose a basis set of wavefunctions having a symmetry of $D^{\frac{1}{2}}$ (Tables S21-S22), in which $D^{\frac{1}{2}} = D^{\frac{1}{2}} \times A = {}^1\overline{E} + {}^2\overline{E}$. According to Step (2), we have:

$$\overline{\overline{D^{\frac{1}{2}} \times D^{\frac{1}{2}}}} = \overline{\overline{({}^1\overline{E} + {}^2\overline{E}) \times ({}^1\overline{E} + {}^2\overline{E})}} = 2A + 2B \qquad \text{(S6)}$$

Then according to the symmetries of $A$ and $B$, we need to derive the symmetry-adapted $k$ terms and C-G coefficients, which are presented in Tables S23 and S24, respectively.

It should also be noted that for $Na_2Sn_2O_3$ and $AgClO_4$, only the in-plane spin components are allowed since all the $k$ points in the $k_x - k_y$ plane have $TC_{2z}$ symmetry. Now, according to Step (4), the Hamiltonian at $\boldsymbol{k}$-point with $C_{2y}$ symmetry, where we consider the spin-orbit coupling terms constrained in the $k_x - k_y$ plane, can be expressed as:

$$H = \beta_1 k_x \sigma_x + \beta_2 k_y \sigma_y \qquad \text{(S7)}$$

Similarly, the Hamiltonian at ***k***-point with $C_{2x}$ symmetry (here, it is the CBM ($k_x$, 0, 0) in $Na_2Sn_2O_3$, Σ and the VBM ($k_x$, 0, 0) in $AgClO_4$) can be obtained:

$$H = \beta_1 k_x \sigma_x + \beta_2 k_y \sigma_y \quad \text{(S8)}$$

**Determination of the eigenvalue and eigenvector of the $\boldsymbol{k} \cdot \boldsymbol{p}$ model**

Our study is interested in the PST region of $Na_2Sn_2O_3$, which is in the $k_y$ path having $C_{2y}$ symmetry. Therefore, we only solve for the eigenvalue and eigenvector of the Hamiltonian having $C_{2y}$ symmetry. We then write the Hamiltonian at the CBM (0, $k_y$, 0) in a matrix form:

$$\mathcal{H} = \mathcal{H}_0 + \begin{vmatrix} \mathcal{H}'_{11} & \mathcal{H}'_{12} \\ {\mathcal{H}'_{12}}^* & -\mathcal{H}'_{11} \end{vmatrix} \quad \text{(S9)}$$

where $\mathcal{H}'_{11} = \beta_2 k_y$ and $\mathcal{H}'_{12} = -\beta_1 k_x i$ and ${\mathcal{H}'_{i2}}^*$ is the conjugate part of $\mathcal{H}'_{i2}$, and $\gamma$ is the real constant. We chose the basis set of the wavefunctions for the two-band as $D^{1/2}$, which represents the wavefunctions comprised of *p* or *d* orbitals of spin 1/2. The energy will be:

$$E = \begin{cases} E_0 - \sqrt{(\beta_1 k_x)^2 + (\beta_2 k_y)^2} \\ E_0 + \sqrt{(\beta_1 k_x)^2 + (\beta_2 k_y)^2} \end{cases} \quad \text{(S10)}$$

where we consider $\beta_2 k_y > 0$. We can obtain the strength of the SOC parameter $\beta$ from the band dispersion relations, and $E_0$ is the energy of $\mathcal{H}_0$ around the CBM.

The wavefunctions are:

$$\psi = \begin{cases} \begin{pmatrix} \frac{\beta_2 k_y - \sqrt{(\beta_1 k_x)^2 + (\beta_2 k_y)^2}}{\beta_1 k_x i} \\ 1 \end{pmatrix} \\ \begin{pmatrix} \frac{\beta_2 k_y + \sqrt{(\beta_1 k_x)^2 + (\beta_2 k_y)^2}}{\beta_1 k_x i} \\ 1 \end{pmatrix} \end{cases} \quad \text{(S11)}$$

When $k_x = 0$:

$$E = \begin{cases} E_0 - \beta_2 k_y \\ E_0 + \beta_2 k_y \end{cases} \text{and } \psi = \begin{cases} \binom{0}{1} \\ \binom{1}{0} \end{cases} \qquad \text{(S12)}$$

The unidirectional spin directions $\langle s_y \rangle$ for the model are -1/2 and 1/2 for the two conduction band states with increasing energy.

When $k_x$ is infinitesimal:

$$E = \begin{cases} E_0 - \beta_2 k_y - \frac{(\beta_1 k_x)^2}{2\beta_2 k_y} \\ E_0 + \beta_2 k_y + \frac{(\beta_1 k_x)^2}{2\beta_2 k_y} \end{cases} \text{and } \psi = \begin{cases} \begin{pmatrix} \frac{\beta_1 k_x i}{2\beta_2 k_y} \\ 1 \end{pmatrix} \\ \begin{pmatrix} 1 \\ \frac{\beta_1 k_x i}{2\beta_2 k_y} \end{pmatrix} \end{cases} \qquad \text{(S13)}$$

And the spin directions are:

$$\begin{cases} \langle s_x \rangle = \frac{<\psi_2|s_y|\psi_2>}{<\psi_2|\psi_2>} \approx \frac{\beta_1 k_x}{2\beta_2 k_y} \\ \langle s_y \rangle = \frac{<\psi_2|s_z|\psi_2>}{<\psi_2|\psi_2>} \approx \frac{1}{2} \\ \langle s_z \rangle = \frac{<\psi_2|s_x|\psi_2>}{<\psi_2|\psi_2>} = 0 \end{cases} \qquad \text{(S14)}$$

where $|\psi_2\rangle$ is the wavefunction of the lowest conduction band state from the model. We find that the expectation values $\langle s_x \rangle$ away from the PST (spin deviations) are proportional to $k_x/k_y$, which indicates that there will be a small spin deviation if the occupied *k* point is away from the CBM (0, $k_y$, 0).

For $AgClO_4$, the PST region is located in the $k_y$ path having $C_{2x}$ symmetry. Therefore, we only solve for the eigenvalue and eigenvector of the Hamiltonian having $C_{2x}$ symmetry. We then write the Hamiltonian at the VBM ($k_x$, 0, 0) in a matrix form:

$$\mathcal{H} = \mathcal{H}_0 + \begin{vmatrix} \mathcal{H}'_{11} & \mathcal{H}'_{12} \\ {\mathcal{H}'_{12}}^* & -\mathcal{H}'_{11} \end{vmatrix} \qquad \text{(S15)}$$

where $\mathcal{H}'_{11} = \beta_1 k_x$ and $\mathcal{H}'_{12} = \beta_2 k_y$ and ${\mathcal{H}'_{i2}}^*$ is the conjugate part of $\mathcal{H}'_{i2}$, and $\gamma$ is the real constant. We chose the basis set of the wavefunctions for the two-band as $D^{1/2}$, which represent the wavefunctions comprised of *p* or *d* orbitals of spin 1/2. The energy will be:

$$E = \begin{cases} E_0 - \sqrt{(\beta_1 k_x)^2 + (\beta_2 k_y)^2} \\ E_0 + \sqrt{(\beta_1 k_x)^2 + (\beta_2 k_y)^2} \end{cases} \quad \text{(S16)}$$

where we consider $\beta_1 k_x > 0$. We can obtain the strength of the SOC $\beta$ quantified from the band dispersion relations, and $E_0$ is the energy of $\mathcal{H}_0$ about the VBM.

The wavefunctions are:

$$\psi = \begin{cases} \begin{pmatrix} \frac{\beta_1 k_x - \sqrt{(\beta_1 k_x)^2 + (\beta_2 k_y)^2}}{\beta_2 k_y} \\ 1 \end{pmatrix} \\ \begin{pmatrix} \frac{\beta_1 k_x + \sqrt{(\beta_1 k_x)^2 + (\beta_2 k_y)^2}}{\beta_2 k_y} \\ 1 \end{pmatrix} \end{cases} \quad \text{(S17)}$$

When $k_y = 0$:

$$E = \begin{cases} E_0 - \beta_1 k_x \\ E_0 + \beta_1 k_x \end{cases} \text{and } \psi = \begin{cases} \binom{0}{1} \\ \binom{1}{0} \end{cases} \quad \text{(S18)}$$

The unidirectional spin directions $\langle s_x \rangle$ for the model are -1/2 and 1/2 for the two valence band states with decreasing energy.

When $k_y$ is infinitesimal:

$$E = \begin{cases} E_0 - \beta_1 k_x - \frac{(\beta_2 k_y)^2}{2\beta_1 k_x} \\ E_0 + \beta_1 k_x + \frac{(\beta_2 k_y)^2}{2\beta_1 k_x} \end{cases} \text{and } \psi = \begin{cases} \begin{pmatrix} \frac{\beta_1 k_x i}{2\beta_2 k_y} \\ 1 \end{pmatrix} \\ \begin{pmatrix} 1 \\ \frac{\beta_1 k_x i}{2\beta_2 k_y} \end{pmatrix} \end{cases} \quad \text{(S19)}$$

And the spin directions are:

$$\begin{cases} \langle s_x \rangle = \frac{<\psi_1|s_z|\psi_1>}{<\psi_1|\psi_1>} \approx -\frac{1}{2} \\ \langle s_y \rangle = \frac{<\psi_1|s_x|\psi_1>}{<\psi_1|\psi_1>} \approx -\frac{\beta_2 k_y}{2\beta_1 k_x} \\ \langle s_z \rangle = \frac{<\psi_1|s_y|\psi_1>}{<\psi_1|\psi_1>} = 0 \end{cases} \quad \text{(S20)}$$

where $|\psi_1\rangle$ is the wavefunction of the highest valence band state from the model. We find that the expectation values $\langle s_y \rangle$ away from the PST (spin deviations) are proportional to $k_y/k_x$, which indicates that there will be a small spin deviation if the occupied $k$ point is away from the VBM ($k_x$, 0, 0).

**Spin diffusion equation**

We first follow in line with a previous study[12] in which the spin-charge density dynamic equation can be expressed as:

$$\partial_t \hat{g} + \nabla_{\boldsymbol{R}} \cdot \left\{\frac{1}{2}\overline{\widehat{\boldsymbol{V}}}, \hat{g}\right\} + i[\overline{\boldsymbol{B}(\boldsymbol{k})} \cdot \widehat{\boldsymbol{\sigma}}, \hat{g}] + \frac{\hat{g}}{\tau} = \frac{\hat{\rho}(\boldsymbol{R},T)}{\tau} \quad \text{(S21)}$$

where $\widehat{\boldsymbol{V}} = \partial\mathcal{H}/\partial\boldsymbol{k}$ and $\boldsymbol{B}(\boldsymbol{k}) = (\beta_1 k_x, \beta_2 k_y, 0)$, $\overline{\widehat{\boldsymbol{V}}}$ and $\overline{\boldsymbol{B}}$ indicates the average quantities; $\tau$ is the momentum scattering time; $\hat{g}$ and $\hat{\rho}$ are thermal average distribution function and density matrix, respectively, which can be written as:

$$\hat{g} = g_c\sigma_0 + g_x\sigma_x + g_y\sigma_y + g_z\sigma_z$$
$$\hat{\rho} = \rho_c\sigma_0 + \rho_x\sigma_x + \rho_y\sigma_y + \rho_z\sigma_z \quad \text{(S22)}$$

The spin lifetime for $\mathcal{H}$ having $C_{2y}$ and $C_{2x}$ symmetry (S7 and S8) can be derived as below:

$$\nabla_{\boldsymbol{R}} \cdot \left\{\frac{1}{2}\overline{\widehat{\boldsymbol{V}}}, \hat{g}\right\} = \frac{\overline{\boldsymbol{k}}}{m} \cdot \nabla_{\boldsymbol{R}}\hat{g} + \beta_2\sigma_y\partial_y g_c + \beta_1\sigma_x\partial_x g_c + (\beta_1\partial_x g_x + \beta_2\partial_y g_y)\sigma_0 \quad \text{(S23)}$$

$$i[\overline{\boldsymbol{B}(\boldsymbol{k})} \cdot \widehat{\boldsymbol{\sigma}}, \hat{g}] = i[\beta_1 k_x\sigma_x + \beta_2 k_y\sigma_y, \hat{g}] = -2\beta_2\overline{k_y}g_z\sigma_x + 2\beta_1\overline{k_x}g_z\sigma_y + 2(\beta_2\overline{k_y}g_x - \beta_1\overline{k_x}g_y)\sigma_z \quad \text{(S24)}$$

where $\overline{\boldsymbol{k}}$ is the thermal average momentum and there is a constant of $\hbar$ in $\overline{\boldsymbol{k}}$ for the simplicity of writing the following equations. Inserting (S23) and (S24) to (S21), we obtain the equations:

$$\left[\left(\partial_t+\frac{1}{\tau}+\frac{\overline{\boldsymbol{k}}}{m}\cdot\boldsymbol{\nabla}_{\boldsymbol{R}}\right)g_c+\beta_1\partial_x g_x+\beta_2\partial_y g_y\right]\sigma_0=\frac{\rho_c\sigma_0}{\tau}$$

$$\left[\left(\partial_t+\frac{1}{\tau}+\frac{\overline{\boldsymbol{k}}}{m}\cdot\boldsymbol{\nabla}_{\boldsymbol{R}}\right)g_x-2\beta_2\overline{k_y}g_z+\beta_1\partial_x g_c\right]\sigma_x=\frac{\rho_x\sigma_x}{\tau}$$

$$\left[\left(\partial_t+\frac{1}{\tau}+\frac{\overline{\boldsymbol{k}}}{m}\cdot\boldsymbol{\nabla}_{\boldsymbol{R}}\right)g_y+2\beta_1\overline{k_x}g_z+\beta_2\partial_y g_c\right]\sigma_y=\frac{\rho_y\sigma_y}{\tau}$$

$$\left[\left(\partial_t+\frac{1}{\tau}+\frac{\overline{\boldsymbol{k}}}{m}\cdot\boldsymbol{\nabla}_{\boldsymbol{R}}\right)g_z+2\left(\beta_2\overline{k_y}g_x-\beta_1\overline{k_x}g_y\right)\right]\sigma_z=\frac{\rho_z\sigma_z}{\tau}$$

(S25)

Writing (S25) into a matrix form and performing the Fourier transformation, we obtain:

$$\begin{bmatrix} 1-i\omega\tau+i\boldsymbol{q}\cdot\boldsymbol{v}\tau & i\beta_1 q_x\tau & i\beta_2 q_y\tau & 0 \\ i\beta_1 q_x\tau & 1-i\omega\tau+i\boldsymbol{q}\cdot\boldsymbol{v}\tau & 0 & -2\beta_2\overline{k_y}\tau \\ i\beta_2 q_y\tau & 0 & 1-i\omega\tau+i\boldsymbol{q}\cdot\boldsymbol{v}\tau & 2\beta_1\overline{k_x}\tau \\ 0 & 2\beta_2\overline{k_y}\tau & -2\beta_1\overline{k_x}\tau & 1-i\omega\tau+i\boldsymbol{q}\cdot\boldsymbol{v}\tau \end{bmatrix}\begin{bmatrix} g_c \\ g_x \\ g_y \\ g_z \end{bmatrix}=\begin{bmatrix} \rho_c \\ \rho_x \\ \rho_y \\ \rho_z \end{bmatrix} \qquad \text{(S26)}$$

If the SOC is strong, then $\beta\bar{k}\tau > 1$, the enhanced spin lifetime frequency $i\omega$ will then be numerically computed using Eq. (S26), when the determinant of $D'-I$ is zero.[12] $I$ is the unit matrix and $D'$ is:

$$\int\frac{d\theta}{2\pi}\begin{bmatrix} 1-i\omega\tau+i\boldsymbol{q}\cdot\boldsymbol{v}\tau & i\beta_1 q_x\tau & i\beta_2 q_y\tau & 0 \\ i\beta_1 q_x\tau & 1-i\omega\tau+i\boldsymbol{q}\cdot\boldsymbol{v}\tau & 0 & -2\beta_2\overline{k_y}\tau \\ i\beta_2 q_y\tau & 0 & 1-i\omega\tau+i\boldsymbol{q}\cdot\boldsymbol{v}\tau & 2\beta_1\overline{k_x}\tau \\ 0 & 2\beta_2\overline{k_y}\tau & -2\beta_1\overline{k_x}\tau & 1-i\omega\tau+i\boldsymbol{q}\cdot\boldsymbol{v}\tau \end{bmatrix}^{-1} \qquad \text{(S27)}$$

Last, we emphasize that our investigated materials are all in the strong SOC regime ($\beta\bar{k}\tau > 1$). It should be noted that, for the comparison of spin lifetime in diverse systems, the Fermi wavelength ($k_F$) is set as the same value which is 0.067 $Å^{-1}$.

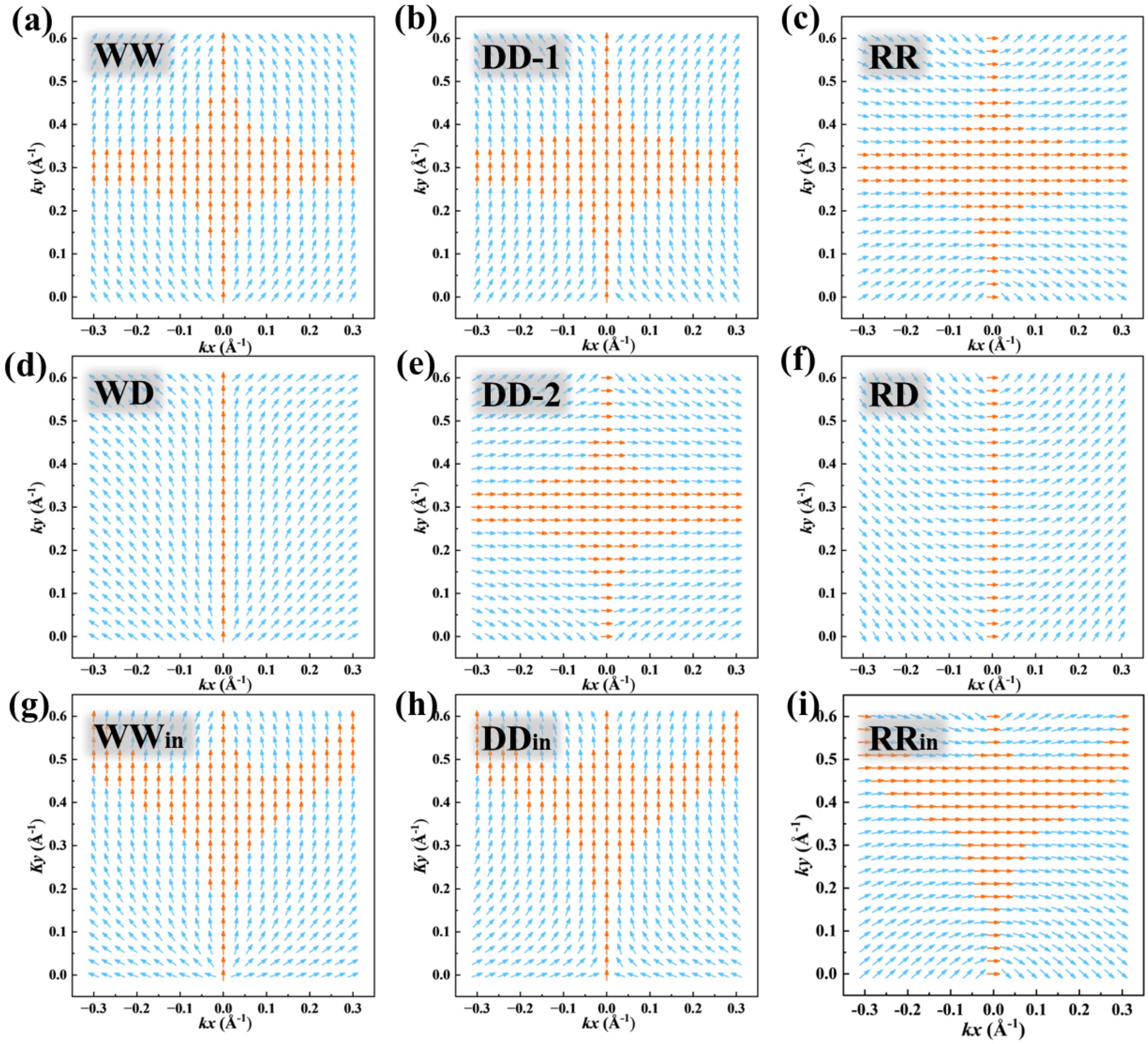

**Figure S1.** Simulation of spin textures based on our models as both spin-orbit fields induce typical spin textures (a-f) and one of the spin-orbit fields is inhomogeneous (g-i). The regions of PST (spin deviation less than 5°) are emphasized with orange arrows, and the rest are labelled with blue arrows.

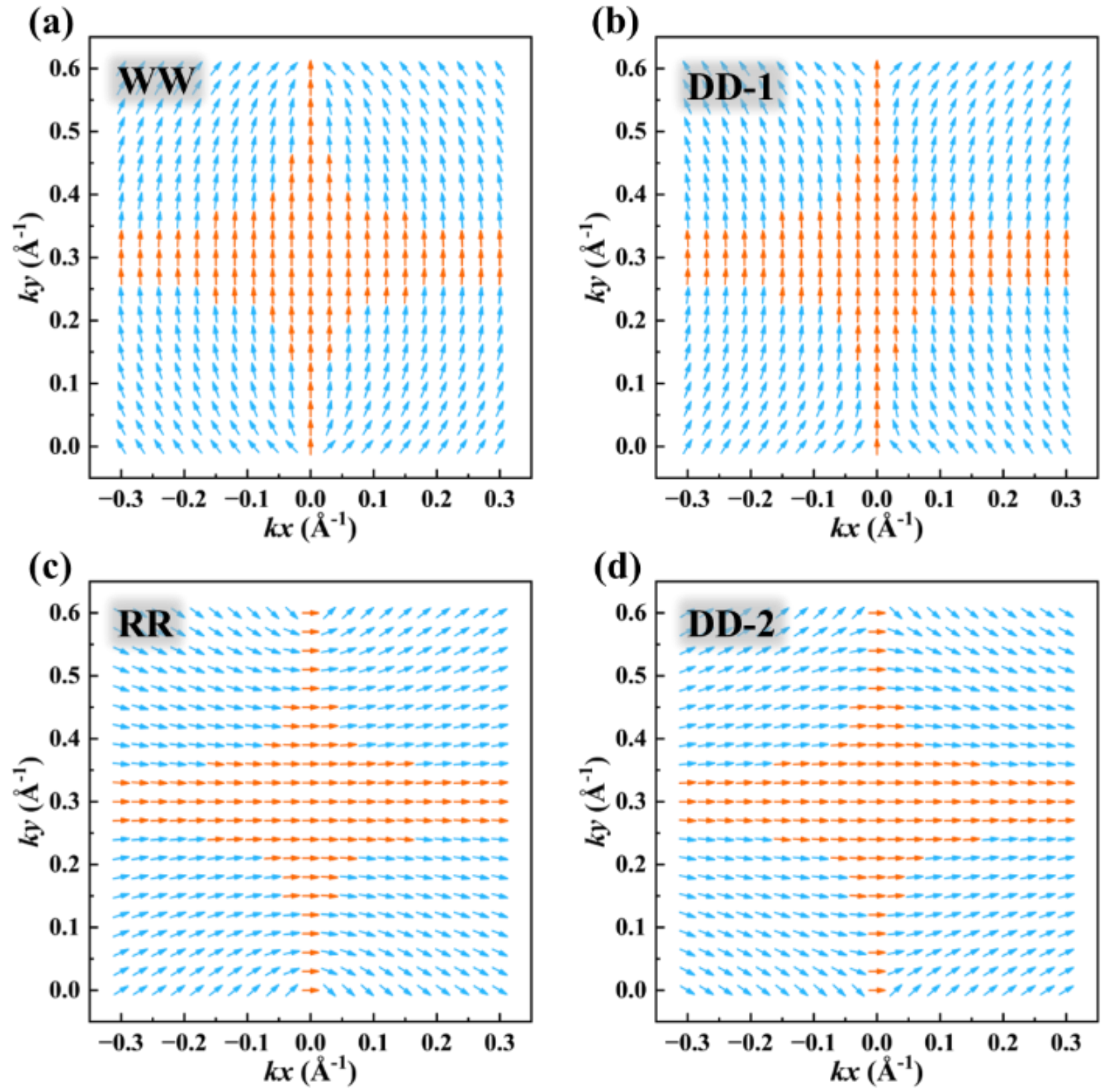

**Figure S2.** Simulation of spin textures based on our models with terms up to third order in $k$ as both spin-orbit fields induce typical spin textures (i.e., $\alpha_i = \beta_i$ and $\alpha_{i1} = \alpha_{i2} = \beta_{i1} = \beta_{i2}$, $i = 1, 2$). The regions of PST (spin deviation less than 5°) are emphasized with orange arrows, and the rest are labelled with blue arrows.

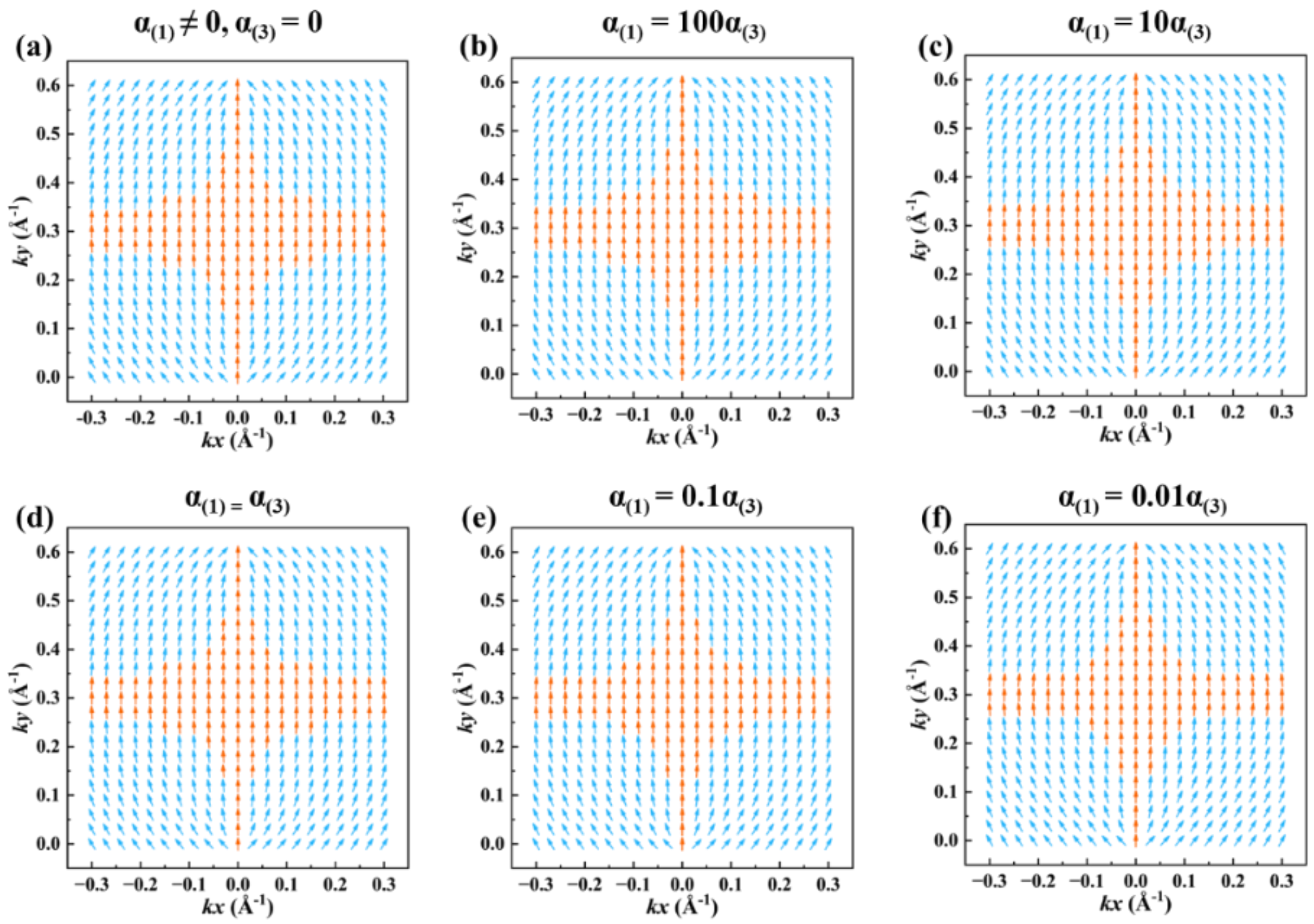

**Figure S3.** Simulation of spin textures based on the WW model including terms up to third order in $k$ with different parameter settings. Both spin-orbit fields induce typical spin textures (i.e., $\alpha_i = \beta_i$ and $\alpha_{i1} = \alpha_{i2} = \beta_{i1} = \beta_{i2}$, $i = 1, 2$). Here $\alpha_{(1)}$ denotes the SOC parameters for linear terms and $\alpha_{(3)}$ denotes those for cubic terms. The regions of PST (spin deviation less than 5°) are emphasized with orange arrows, and the rest are labelled with blue arrows.

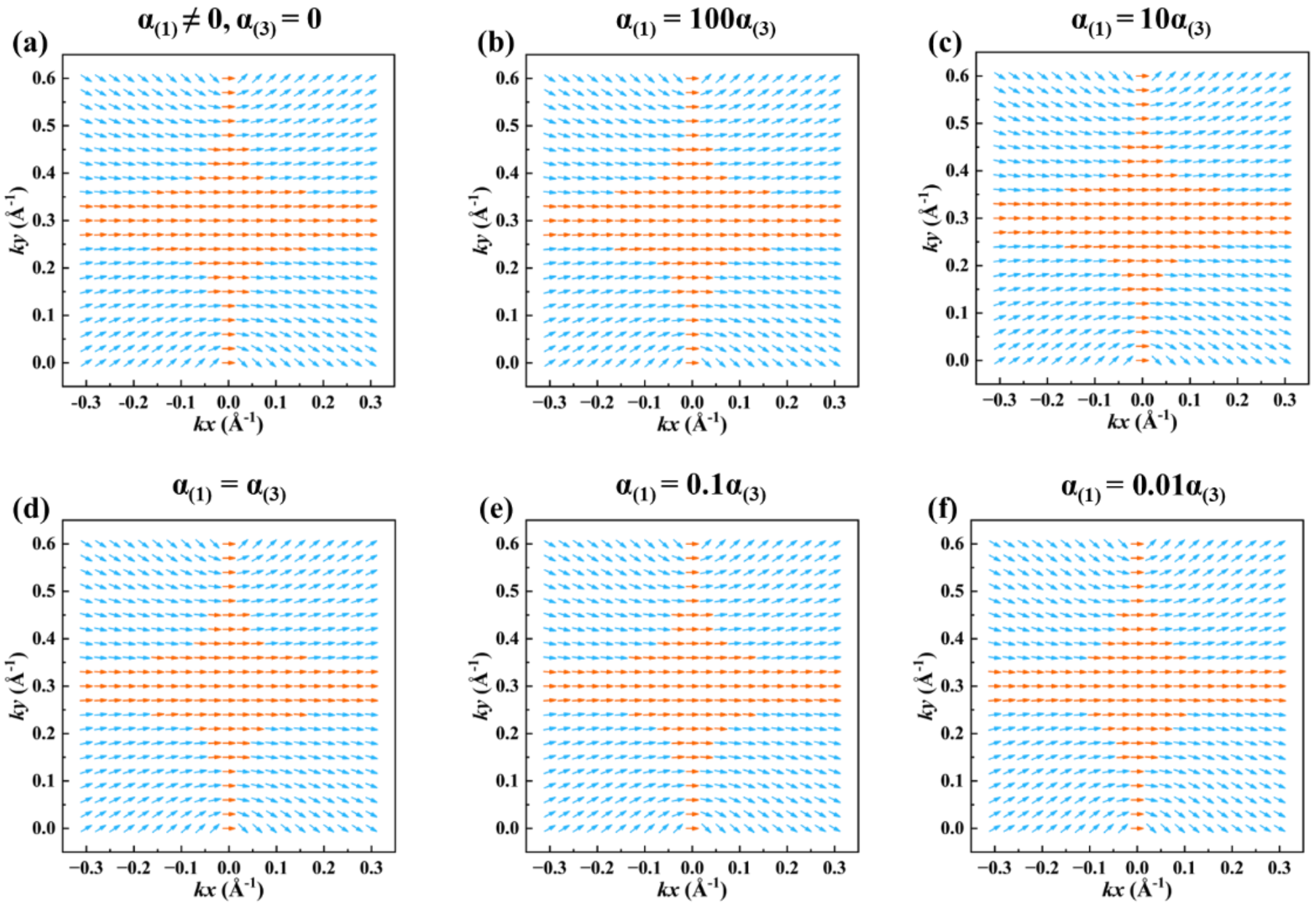

**Figure S4.** Simulation of spin textures based on the RR model including terms up to third order in $k$ with different parameter settings. Both spin-orbit fields induce typical spin textures (i.e., $\alpha_i = \beta_i$ and $\alpha_{i1} = \alpha_{i2} = \beta_{i1} = \beta_{i2}$, $i = 1, 2$). Here $\alpha_{(1)}$ denotes the SOC parameters for linear terms and $\alpha_{(3)}$ denotes those for cubic terms. The regions of PST (spin deviation less than 5°) are emphasized with orange arrows, and the rest are labelled with blue arrows.

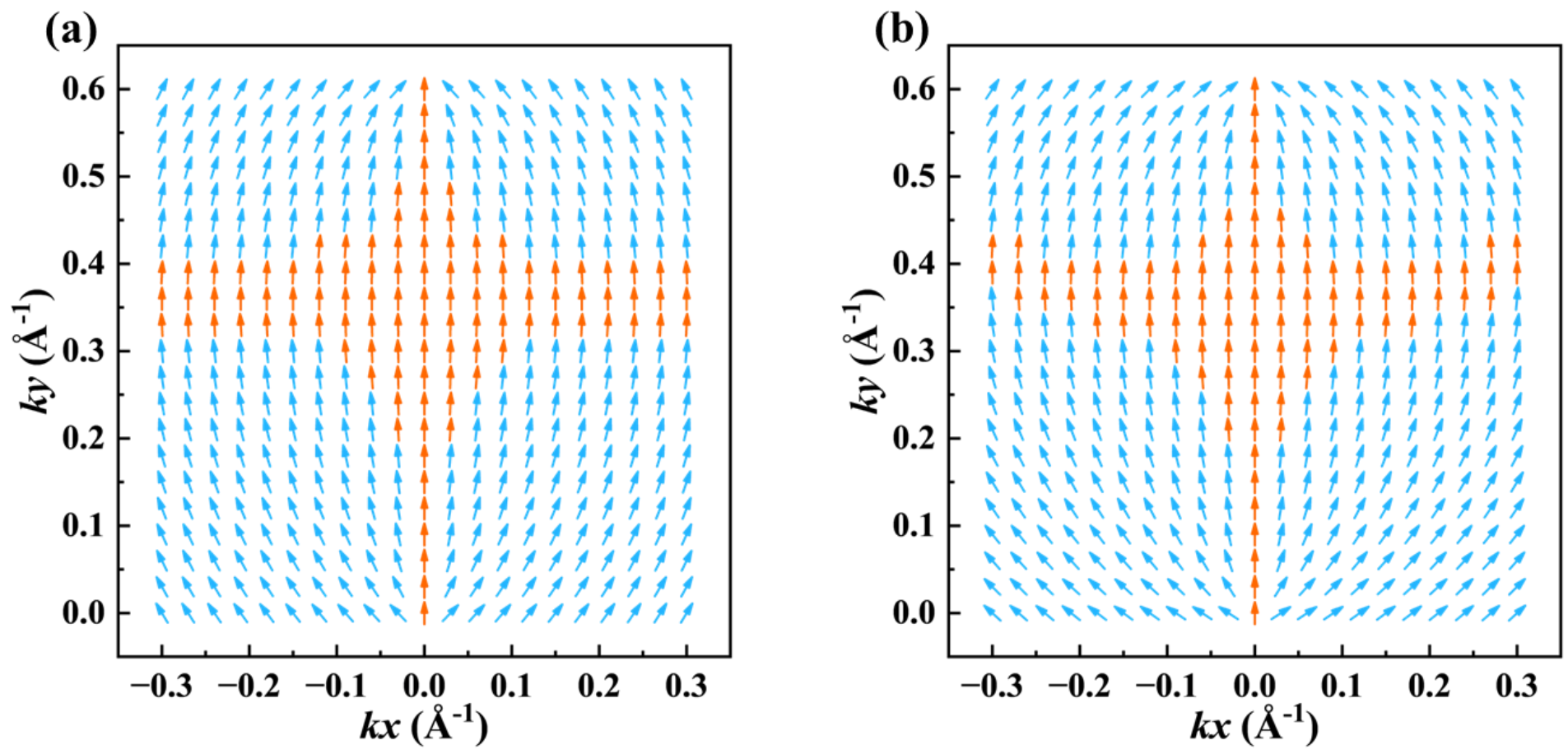

**Figure S5.** Simulation of spin textures based on the WW model with terms up to third order in $k$. Here one of spin-orbit fields induce typical spin textures (i.e., $\alpha_2 = \beta_2$ and $\alpha_{21} = \alpha_{22} = \beta_{21} = \beta_{22}$) and another spin-orbit field is inhomogeneous (panel a: $\alpha_1 \neq \beta_1$ and $\alpha_{11} = \alpha_{12} = \beta_{11} = \beta_{12}$; panel b: $\alpha_1 = \beta_1$, $\alpha_{11} = \beta_{11}$ and $\alpha_{12} = \beta_{12}$). The regions of PST (spin deviation less than 5°) are emphasized with orange arrows, and the rest are marked with blue arrows.

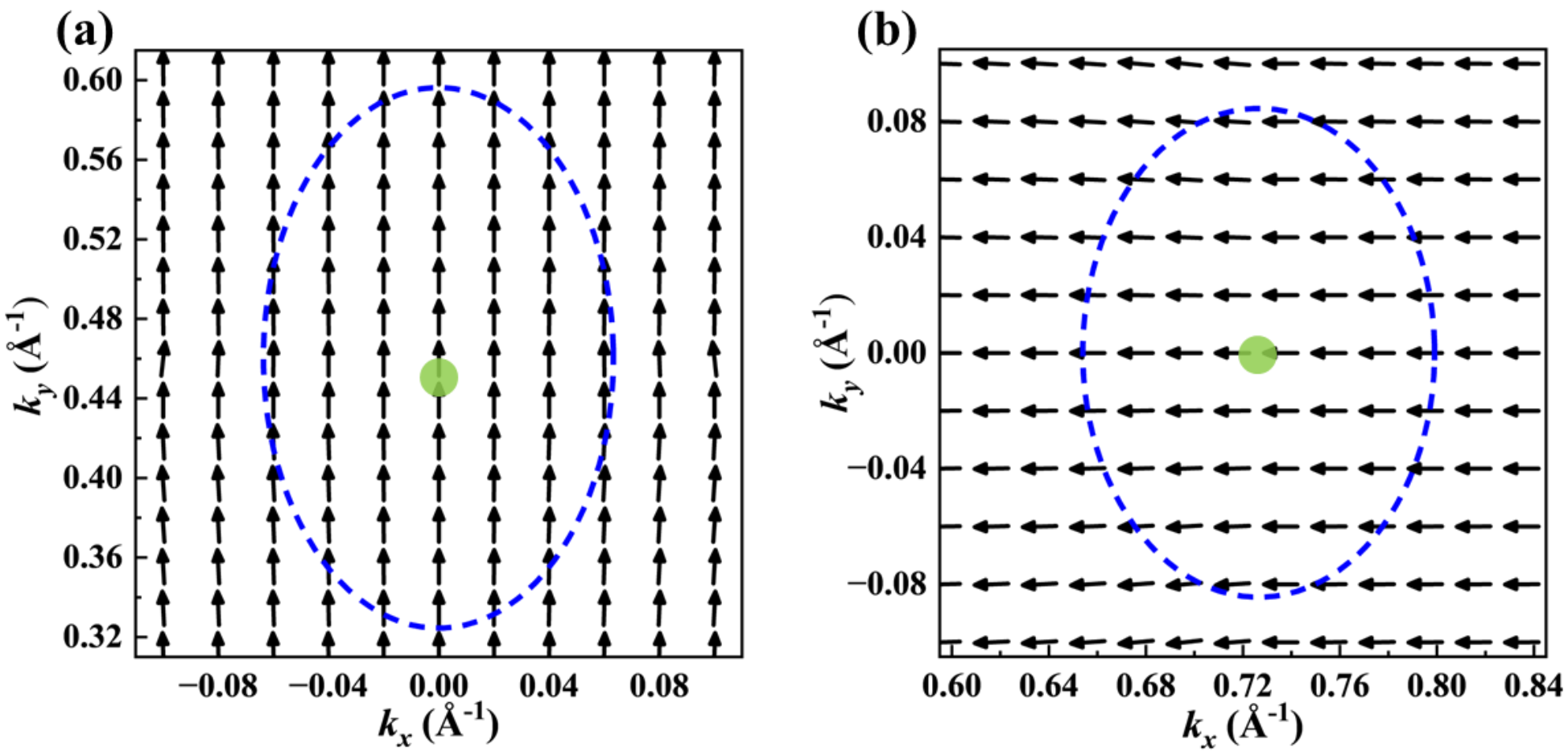

**Figure S6.** Spin textures based on the $k{\cdot}p$ models including terms up to third order in $k$ around (a) the CBM of $Na_2Sn_2O_3$ and (b) the VBM of $AgClO_4$. The high-quality PST regions for carrier doping are highlighted with the blue dashed circle. The location of the CBM/VBM is indicated by the green circle.

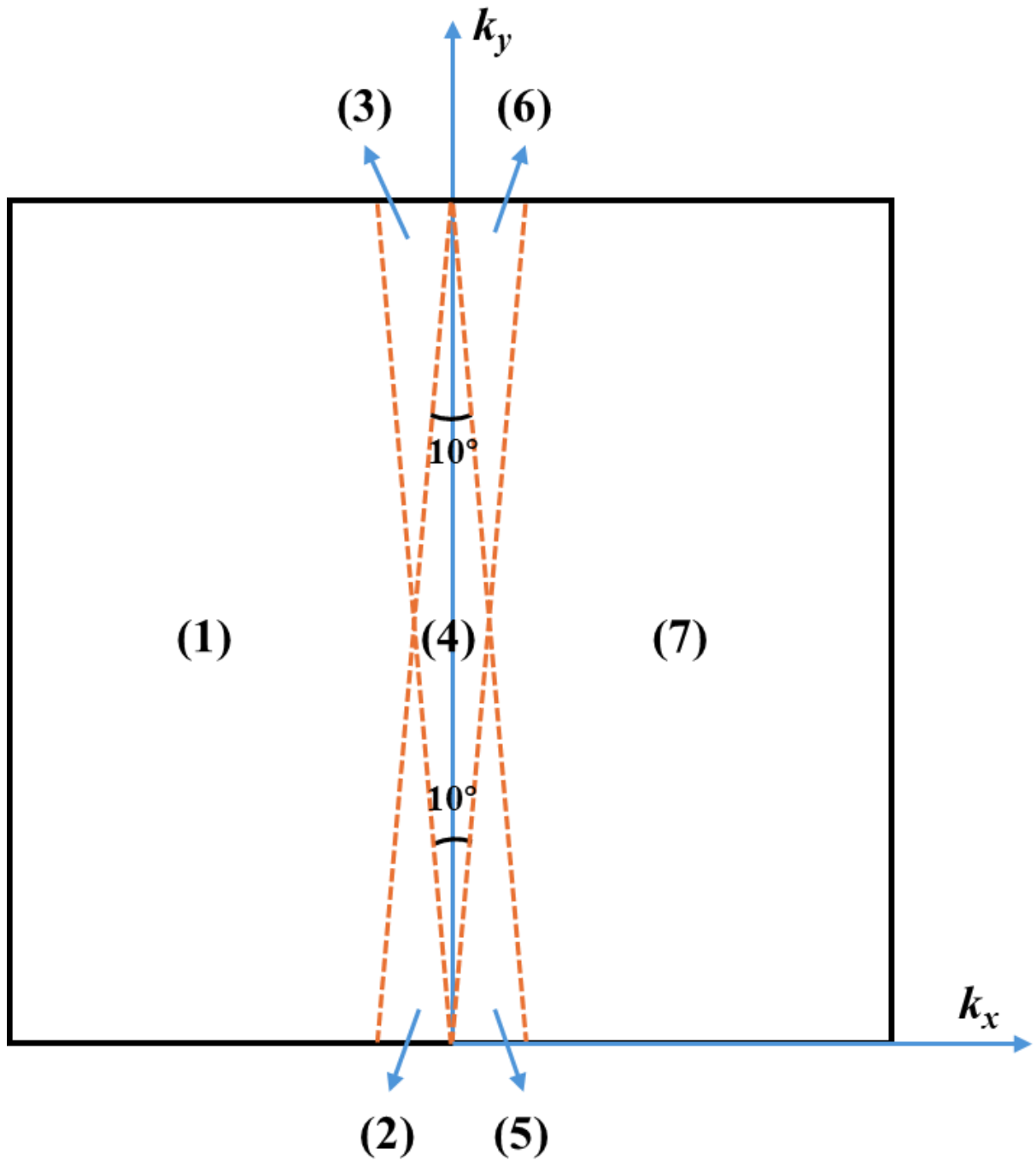

**Figure S7**. Schematic of seven different districts in the momentum space. Seven districts are separated by orange dashed lines.

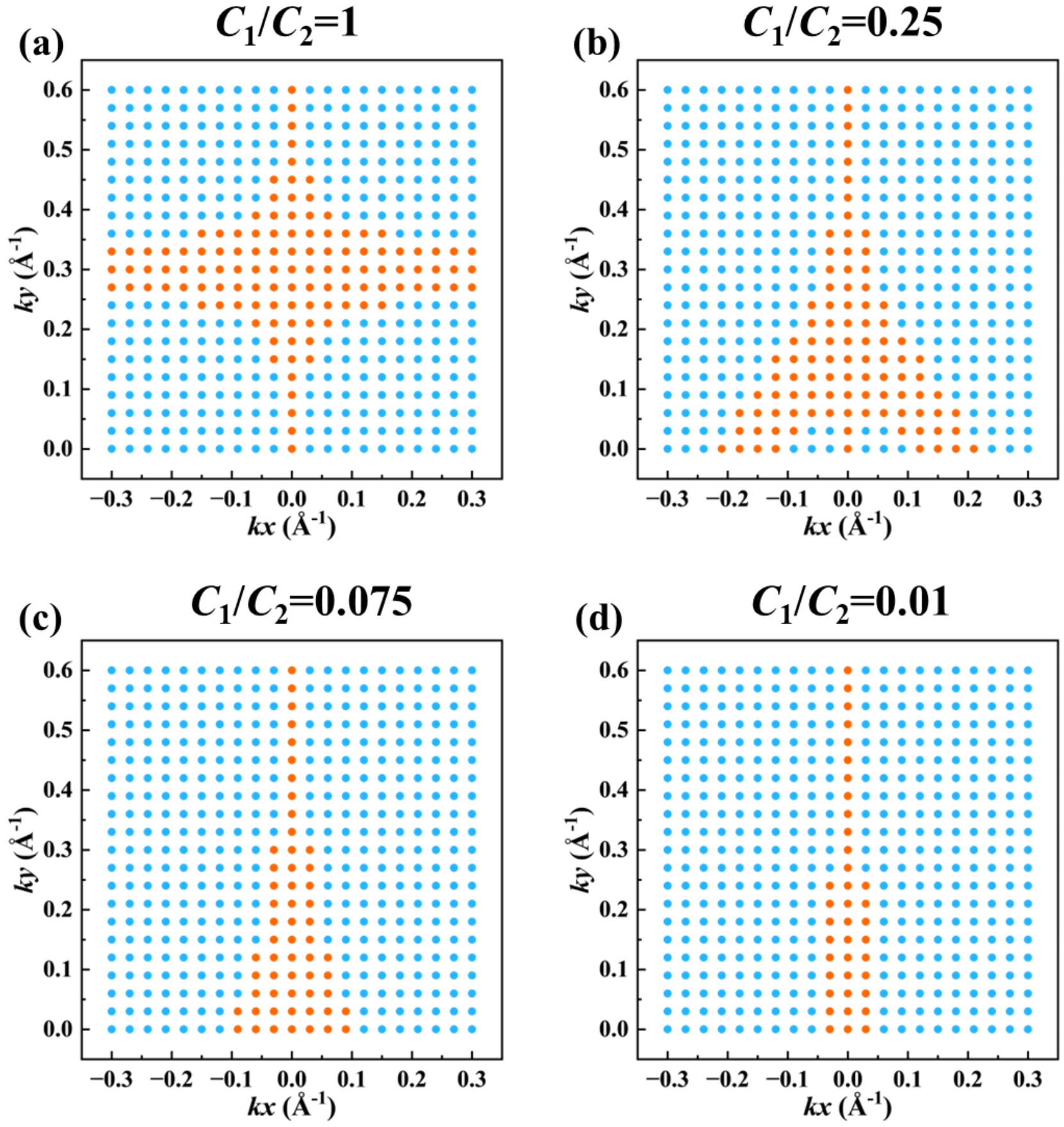

**Figure S8**. Simulation of PST regions with spin deviation less than 5° for different values of $C_1/C_2$ in the WW/DD-2 models. The region of PST is highlighted in orange and the rest of the area is marked in blue.

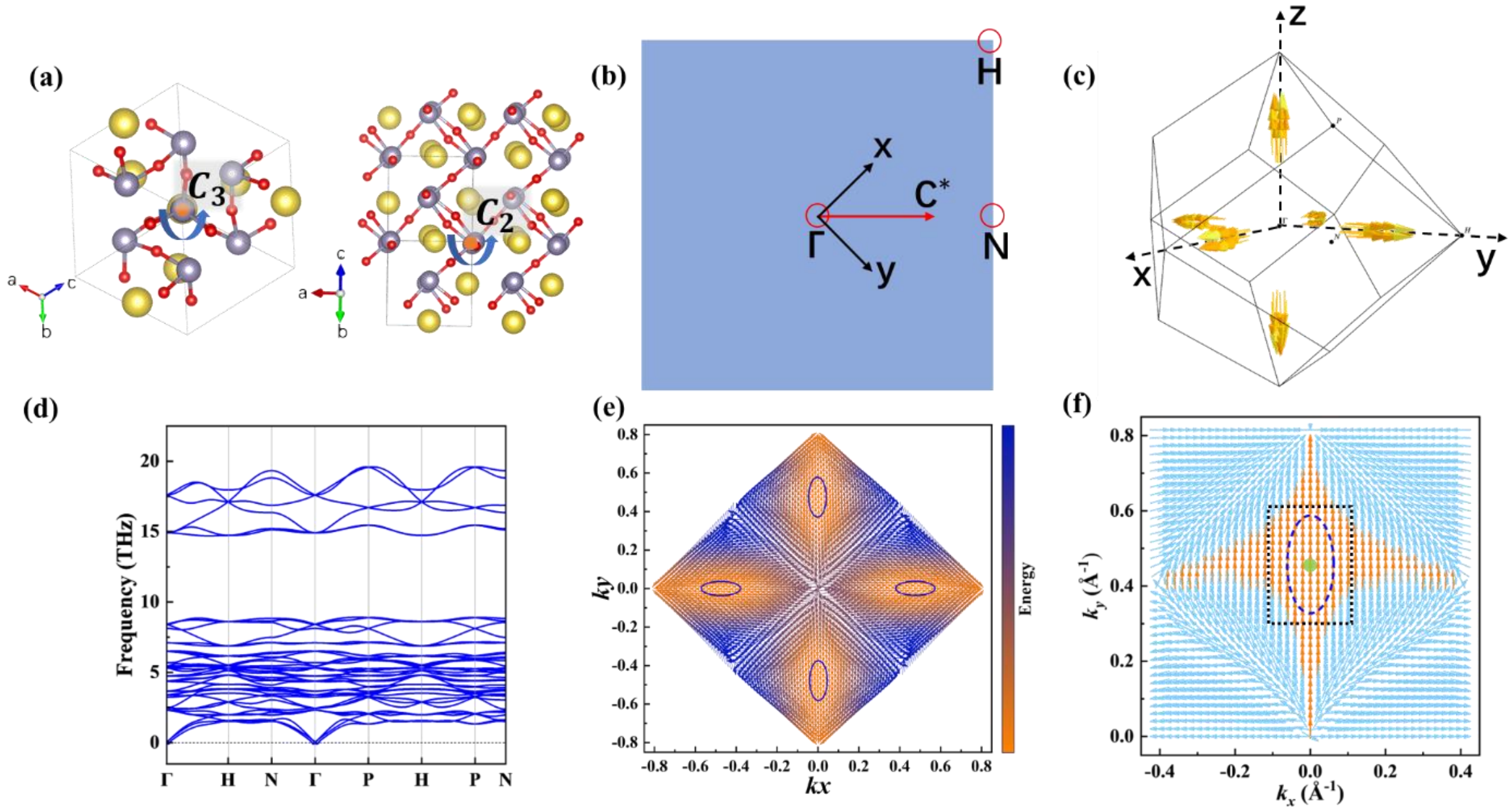

**Figure S9.** (a) Crystal structure and (b) Brillouin zone in the $k_x$-$k_y$ plane of $Na_2Sn_2O_3$. The Na, Sn and O atoms are indicated with yellow, grey and red balls, respectively. (c) Spin texture corresponding to the Fermi surface 48 meV above the CBM in the 3D momentum space. (d) The calculated phonon spectrum of $Na_2Sn_2O_3$. (e) Spin texture in the $k_x$-$k_y$ plane with the region of high-quality PST enclosed by the blue circle. The color represents the projection of energy. (f) Enlargement of the spin texture with the PST about the CBM. PST regions, defined by spin deviation less than 5°, are indicated with orange arrows while the remaining region is marked with blue arrows. Position of the CBM is indicated by green circle. The spin texture enclosed by the black dashed line is displayed in Figure 3b of the main text.

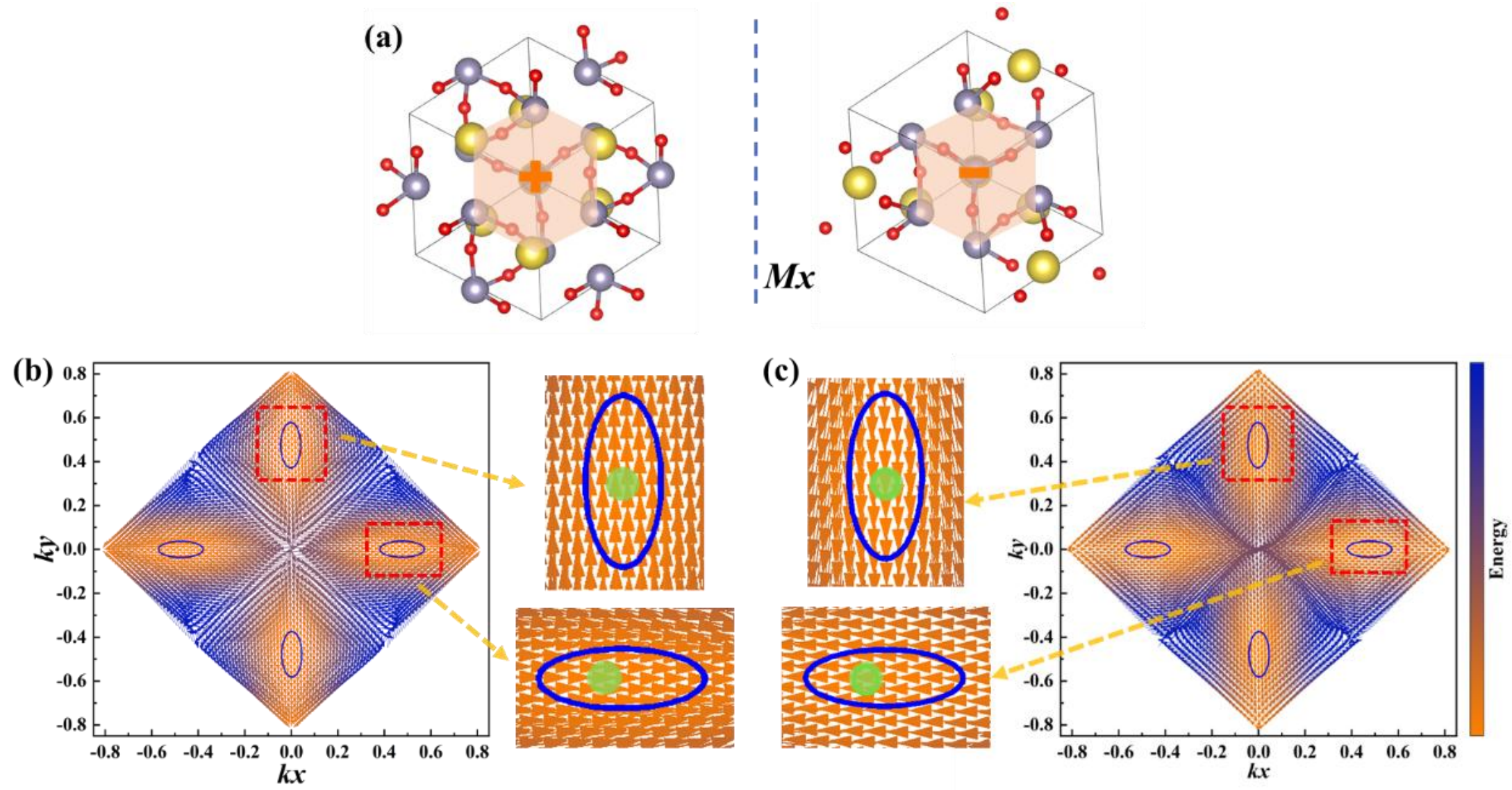

**Figure S10.** (a) The enantiomeric crystals of $Na_2Sn_2O_3$ connected by the mirror symmetry $M_x$. (b, c) Spin textures in the $k_x$-$k_y$ plane and corresponding enlargement of PST with opposite geometric chirality.

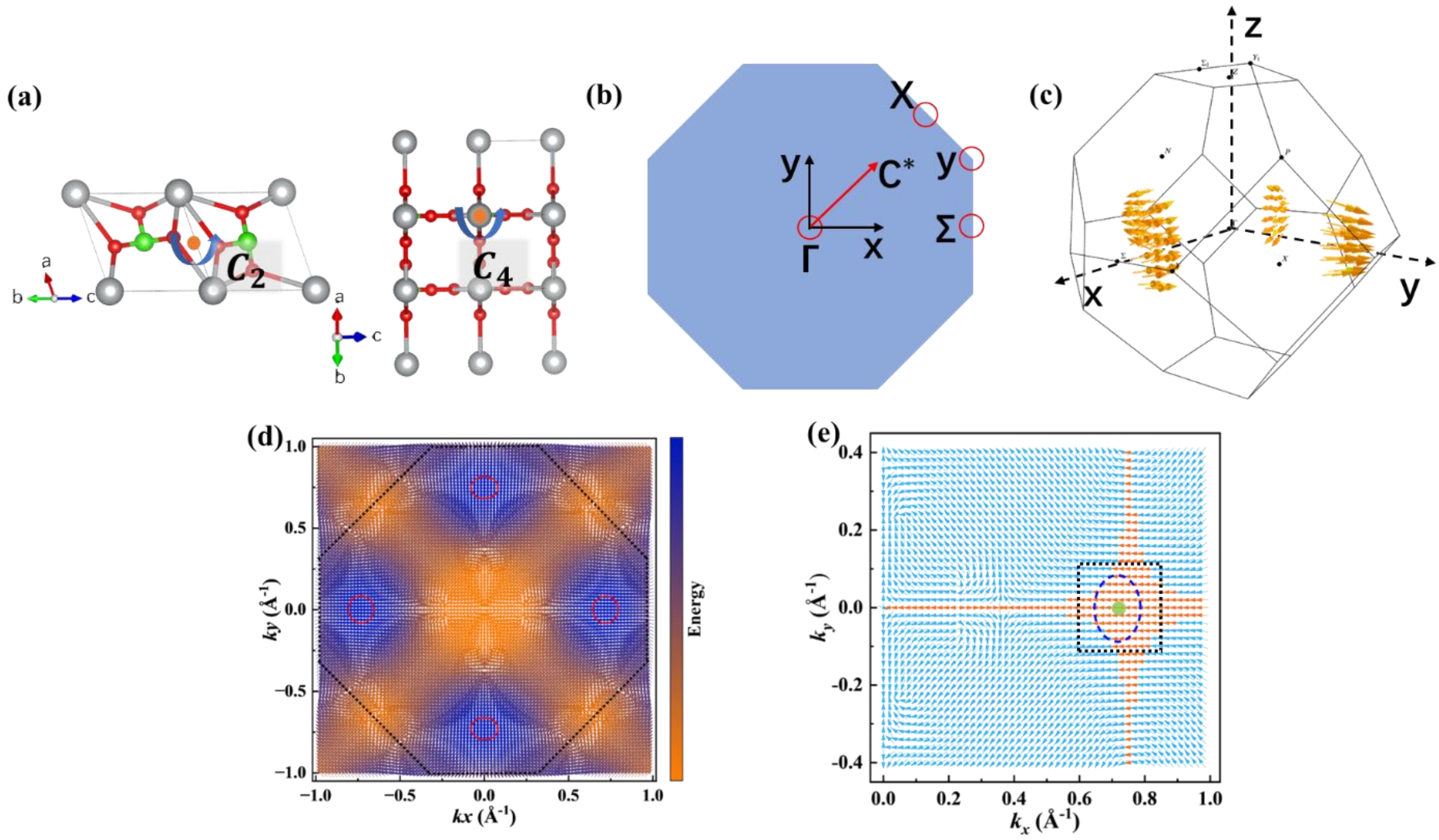

**Figure S11.** (a) Crystal structure and (b) Brillouin zone in the $k_x$-$k_y$ plane of $AgClO_4$. The Ag, Cl and O atoms are indicated with grey, green and red balls, respectively. (c) Spin texture corresponding to the Fermi surface 32 meV below the VBM in the 3D momentum space. (d) Spin texture in the in $k_x$-$k_y$ plane with the region of high-quality PST enclosed by the red circle. The color represents the projection of energy. (e) Enlargement of the spin texture with the PST about the VBM. PST regions, defined by spin deviation less than 5°, are indicated with orange arrows while the remaining region is marked with blue arrows. Position of the VBM is indicated by green circle. The spin texture enclosed by the black dashed line is displayed in Figure 3e of the main text.

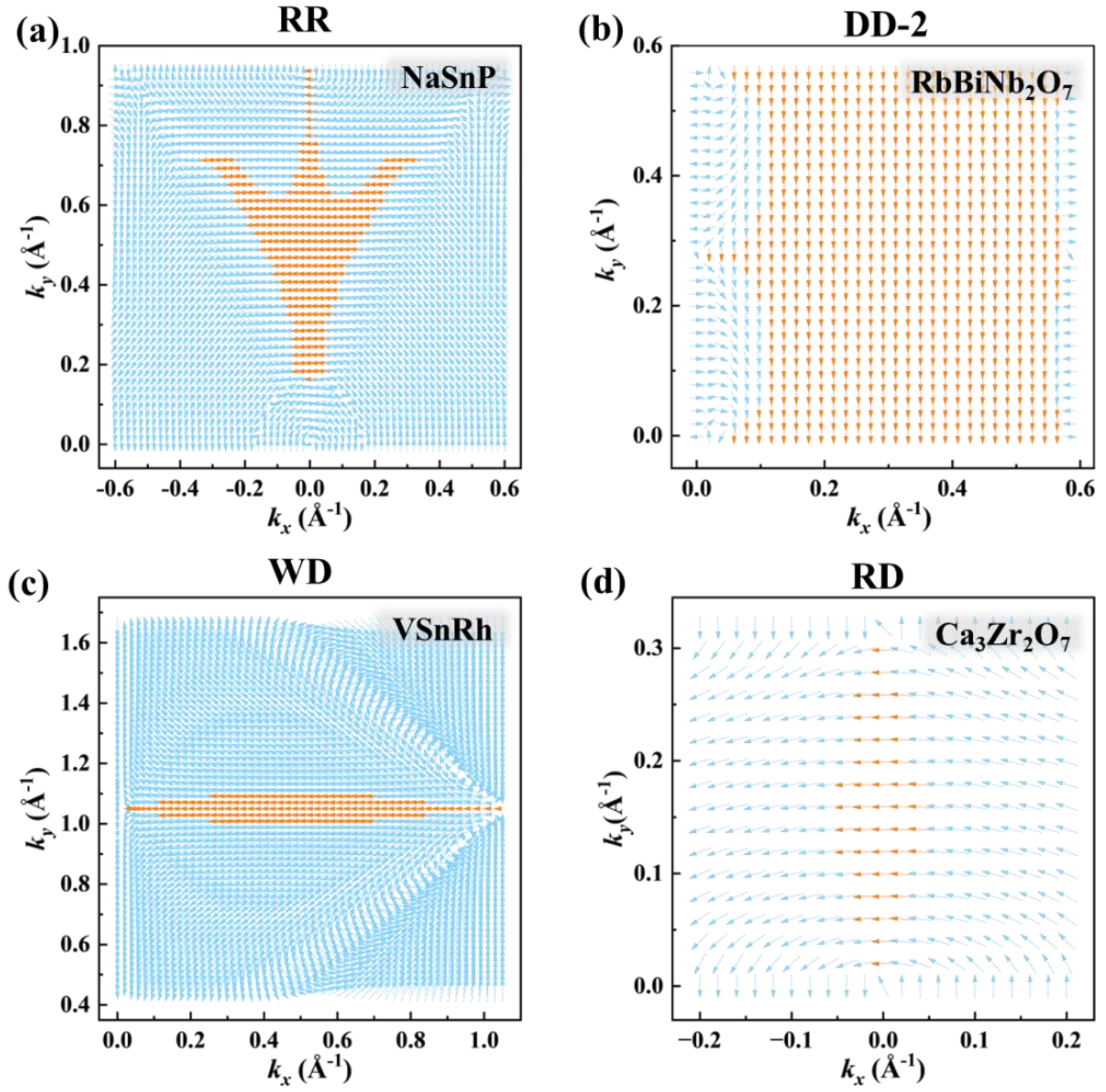

**Figure S12.** Spin textures of some compounds from DFT calculations, corresponding to our four models: (a) RR, (b) DD-2, (c) WD, and (d) RD models.[3–5,13] PST regions with spin deviation less than 5° are indicated with orange arrows, while the remaining regions are marked with blue arrows.

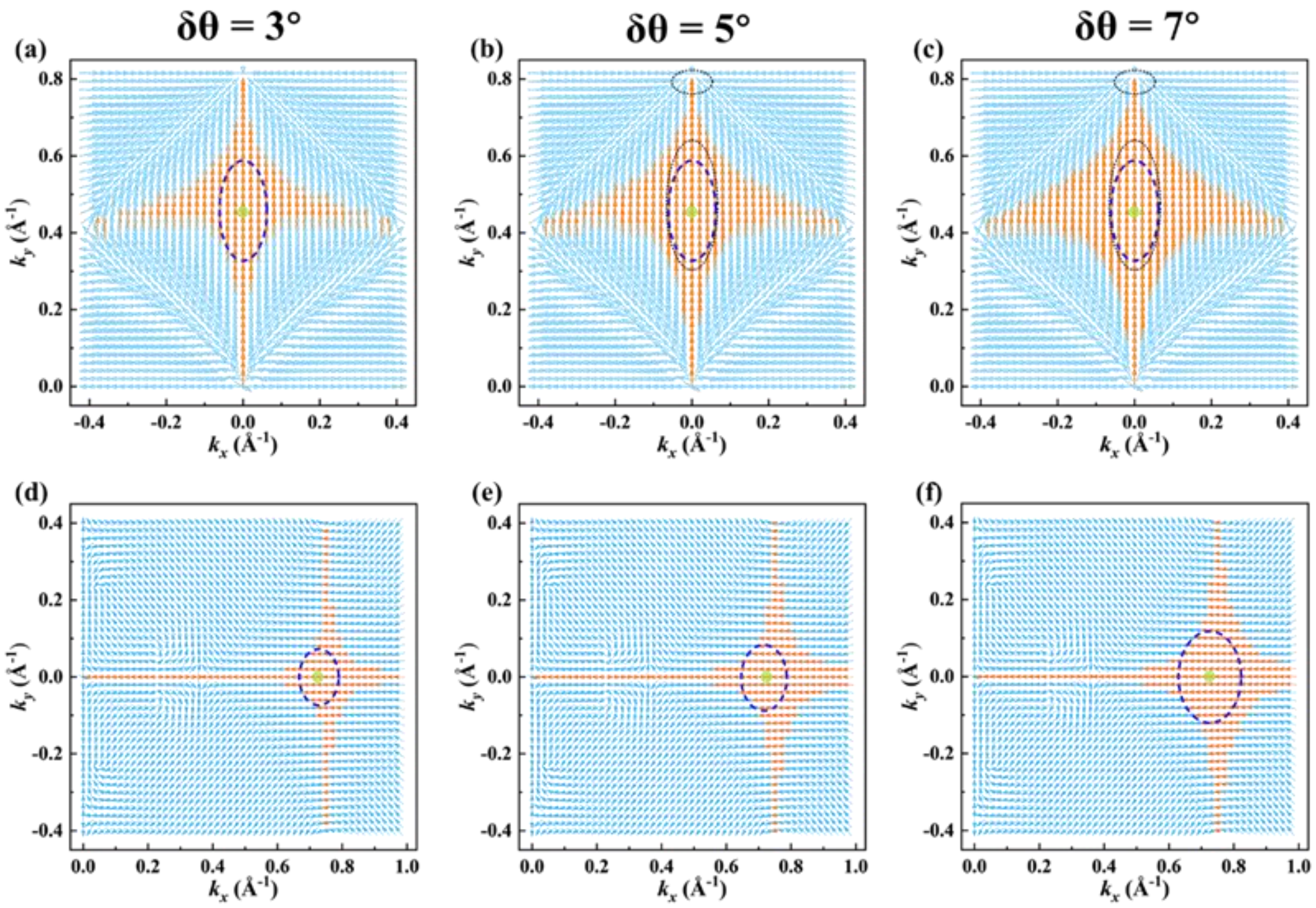

**Figure S13.** The spin textures for (a-c) $Na_2Sn_2O_3$ and (d-f) $AgClO_4$. PST regions with spin deviation δθ less than 3°, 5° and 7° are indicated by orange arrows, while the remaining regions are marked with blue arrows. The location of the CBM is indicated by the green circle and high-quality PST regions for carrier doping are highlighted with the blue dashed circle. In panels (b) and (c), the regions for a higher doping level are enclosed by the black dotted circle.

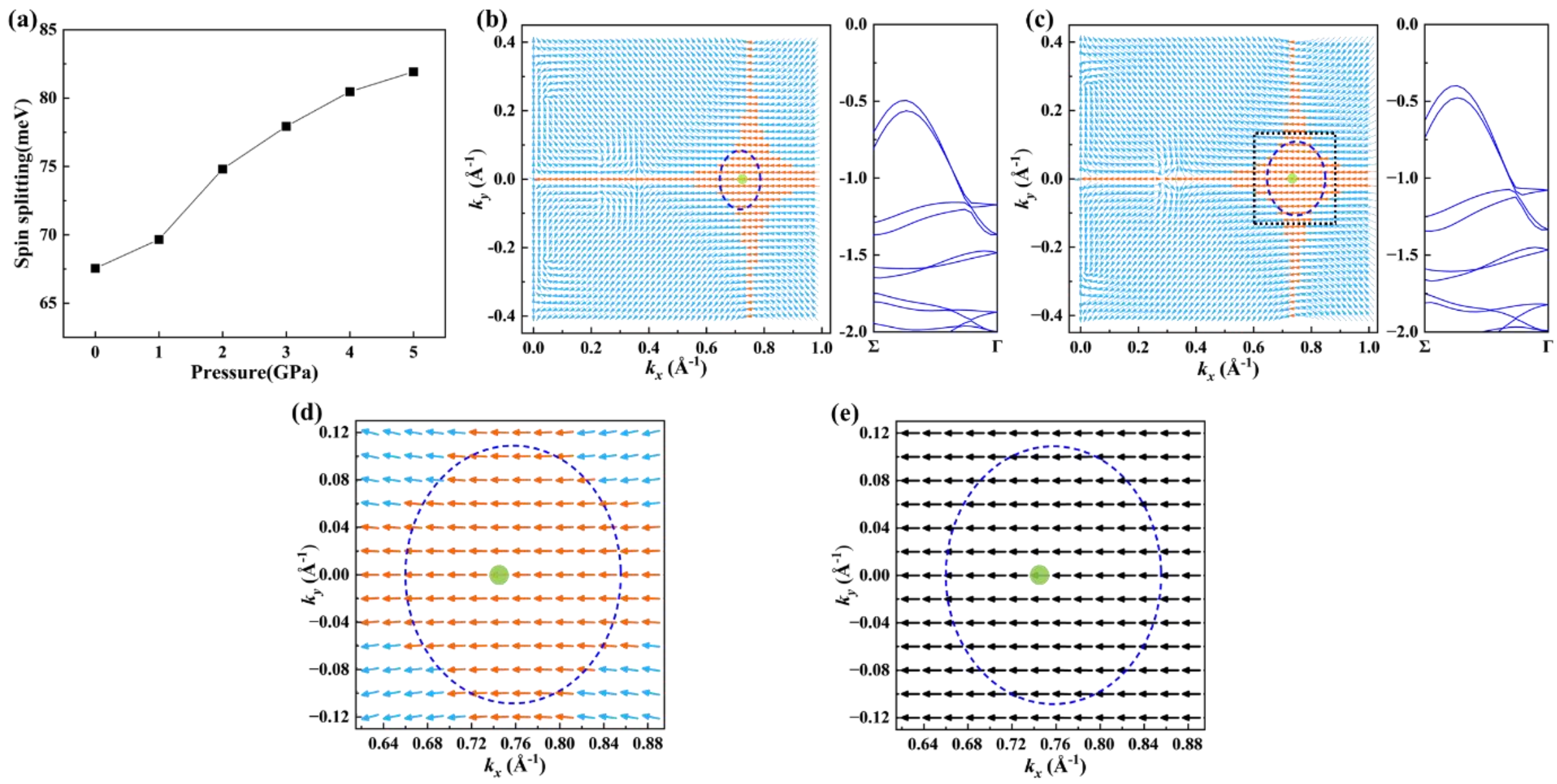

**Figure S14.** (a) The variation curve of spin splitting with applied pressure. (b, c) Spin textures and band structures of $AgClO_4$ (b) before and (c) after applying pressure of 3 GPa. (d) Enlargement of the spin texture enclosed by the black dashed line in panel (c). (e) The spin texture based on the $k{\cdot}p$ model around VBM in $AgClO_4$ under pressure of 3 GPa. PST regions with spin deviation less than 5° are indicated with orange arrows, while the remaining regions are marked with blue arrows. The high-quality PST regions for carrier doping are enclosed by the blue dashed circle and the VBM is labelled with the green circle.

**TABLE S1.** The linear and cubic SOC terms allowed in 11 wavevector point groups.

| Wavevector point group | Linear terms | Cubic terms |
|---|---|---|
| $C_{2v}$ | $k_x\sigma_y$, $k_y\sigma_x$ (RD) | $k_x^3\sigma_y$, $k_x^2k_y\sigma_x$, $k_xk_y^2\sigma_y$, $k_y^3\sigma_x$ |
| $C_{3v}$ | $k_x\sigma_y - k_y\sigma_x$ (R) | $(k_x^3 + k_xk_y^2)\sigma_y - (k_x^2k_y + k_y^3)\sigma_x$,<br>$(k_x^3 - 3k_xk_y^2)\sigma_z$ |
| $C_{4v}$ | $k_x\sigma_y - k_y\sigma_x$ (R) | $k_x^3\sigma_y - k_y^3\sigma_x$, $k_x^2k_y\sigma_x - k_xk_y^2\sigma_y$ |
| $C_{6v}$ | $k_x\sigma_y - k_y\sigma_x$ (R) | $(k_x^3 + k_xk_y^2)\sigma_y - (k_x^2k_y + k_y^3)\sigma_x$ |
| $D_2$ | $k_\alpha\sigma_\alpha$ ($\alpha = x, y$) (WD) | $k_x^3\sigma_x$, $k_xk_y^2\sigma_x$, $k_y^3\sigma_y$, $k_yk_x^2\sigma_y$ |
| $D_3$ | $k_x\sigma_x + k_y\sigma_y$ (W) | $(k_x^3 + k_xk_y^2)\sigma_x + (k_y^3 + k_yk_x^2)\sigma_y$,<br>$(k_x^3 - 3k_xk_y^2)\sigma_z$ |
| $D_4$ | $k_x\sigma_x + k_y\sigma_y$ (W) | $k_x^3\sigma_x + k_y^3\sigma_y$, $k_xk_y^2\sigma_x + k_yk_x^2\sigma_y$ |
| $D_6$ | $k_x\sigma_x + k_y\sigma_y$ (W) | $(k_x^3 + k_xk_y^2)\sigma_x + (k_y^3 + k_yk_x^2)\sigma_y$ |
| $T$ | $k_x\sigma_x + k_y\sigma_y$ (W) | $k_x^3\sigma_x + k_y^3\sigma_y$, $k_xk_y^2\sigma_x$, $k_yk_x^2\sigma_y$ |
| $O$ | $k_x\sigma_x + k_y\sigma_y$ (W) | $k_x^3\sigma_x + k_y^3\sigma_y$, $k_xk_y^2\sigma_x + k_yk_x^2\sigma_y$ |
| $D_{2d}$ | $k_x\sigma_x - k_y\sigma_y$ (D) | $(k_x^3 + k_xk_y^2)\sigma_x - (k_x^2k_y + k_y^3)\sigma_y$ |

**TABLE S2.** $k{\cdot}p$ effective Hamiltonians including linear and cubic terms for our models and the corresponding $S_{\mathrm{dv}}/S_{\mathrm{PST}}$. $\alpha_i, \beta_i, \alpha_{i1}, \alpha_{i2}, \beta_{i1}, \beta_{i2}$ $(i = 1, 2)$ denote the SOC coefficients for different spin-orbit fields and the $\sigma_x$, $\sigma_y$ and $\sigma_z$ represent the Pauli matrices. $S_{\mathrm{dv}}$ and $S_{\mathrm{PST}}$ represent the expectation values of spin dominating spin deviation and PST, respectively.

| Type | Model |
|---|---|
| **WW** | $(\alpha_1-\alpha_2)k_x\sigma_x + (\beta_1 k_y + \beta_2 k'_y)\sigma_y + [(\alpha_{12}{k_y}^2 + \alpha_{11}{k_x}^2) - (\alpha_{22}{k'_y}^2 + \alpha_{21}{k_x}^2)]k_x\sigma_x + [(\beta_{11}{k_x}^2 k_y + \beta_{12}{k_y}^3) + (\beta_{21}{k_x}^2 k'_y + \beta_{22}{k'_y}^3)]\sigma_y$ |
| | $\frac{S_{dv}}{S_{PST}} = \frac{(\alpha_1-\alpha_2)k_x + [(\alpha_{12}{k_y}^2 + \alpha_{11}{k_x}^2) - (\alpha_{22}{k'_y}^2 + \alpha_{21}{k_x}^2)]k_x}{\beta_1 k_y + \beta_2 k'_y + \beta_{12}{k_y}^3 + \beta_{22}{k'_y}^3}$ |
| **DD-1** | $(\alpha_2-\alpha_1)k_x\sigma_x + (\beta_1 k_y + \beta_2 k'_y)\sigma_y + [(\alpha_{21}{k_x}^2 + \alpha_{22}{k'_y}^2) - (\alpha_{11}{k_x}^2 + \alpha_{12}{k_y}^2)]k_x\sigma_x + \left[\left(\beta_{12}{k_y}^3 + \beta_{11}{k_x}^2 k_y\right) + (\beta_{22}{k'_y}^3 + \beta_{21}{k_x}^2 k'_y)\right]\sigma_y$ |
| | $\frac{S_{dv}}{S_{PST}} = \frac{(\alpha_2-\alpha_1)k_x + [(\alpha_{21}{k_x}^2 + \alpha_{22}{k'_y}^2) - (\alpha_{11}{k_x}^2 + \alpha_{12}{k_y}^2)]k_x}{\beta_1 k_y + \beta_2 k'_y + \beta_{12}{k_y}^3 + \beta_{22}{k'_y}^3}$ |
| **RR** | $(\beta_1 k_y + \beta_2 k'_y)\sigma_x + (\alpha_2-\alpha_1)k_x\sigma_y + \left[\left(\beta_{12}{k_y}^3 + \beta_{11}{k_x}^2 k_y\right) + (\beta_{22}{k'_y}^3 + \beta_{21}{k_x}^2 k'_y)\right]\sigma_x + [(\alpha_{21}{k_x}^2 + \alpha_{22}{k'_y}^2) - (\alpha_{11}{k_x}^2 + \alpha_{12}{k_y}^2)]k_x\sigma_y$ |
| | $\frac{S_{dv}}{S_{PST}} = \frac{(\alpha_2-\alpha_1)k_x + [(\alpha_{21}{k_x}^2 + \alpha_{22}{k'_y}^2) - (\alpha_{11}{k_x}^2 + \alpha_{12}{k_y}^2)]k_x}{\beta_1 k_y + \beta_2 k'_y + \beta_{12}{k_y}^3 + \beta_{22}{k'_y}^3}$ |
| **DD-2** | $(\beta_1 k_y + \beta_2 k'_y)\sigma_x + (\alpha_1-\alpha_2)k_x\sigma_y + [(\beta_{11}{k_x}^2 k_y + \beta_{12}{k_y}^3) + (\beta_{21}{k_x}^2 k'_y + \beta_{22}{k'_y}^3)]\sigma_x + [(\alpha_{11}{k_x}^2 + \alpha_{12}{k_y}^2) - (\alpha_{21}{k_x}^2 + \alpha_{22}{k'_y}^2)]k_x\sigma_y$ |
| | $\frac{S_{dv}}{S_{PST}} = \frac{(\alpha_1-\alpha_2)k_x + [(\alpha_{11}{k_x}^2 + \alpha_{12}{k_y}^2) - (\alpha_{21}{k_x}^2 + \alpha_{22}{k'_y}^2)]k_x}{\beta_1 k_y + \beta_2 k'_y + \beta_{12}{k_y}^3 + \beta_{22}{k'_y}^3}$ |

**TABLE S3.** $k{\cdot}p$ models including terms up to third order in $k$ and corresponding SOC parameters for the CBM in $Na_2Sn_2O_3$ and the VBM in $AgClO_4$. $\beta_i$ ($i = 1\sim6$) are the SOC coefficients for different terms. $k_{y-c}$ represents the $k_y$ component of the location of CBM in the $k_x$-$k_y$ plane. $k_{x-v}$ represents the $k_x$ component of the location of VBM in the $k_x$-$k_y$ plane.

| materials | ***k·p* model and SOC coefficients** | | | | | | |
|---|---|---|---|---|---|---|---|
| $Na_2Sn_2O_3$ | $H = \beta_1 k_x\sigma_x + \beta_2 k_y\sigma_y + \beta_3 {k_x}^3\sigma_x + \beta_4 k_x {k_y}^2\sigma_x + \beta_5 k_y {k_x}^2\sigma_y + \beta_6 {k_y}^3\sigma_y$ <br> $\langle s_x\rangle = \frac{<\psi_2\|s_y\|\psi_2>}{<\psi_2\|\psi_2>} \approx \frac{\beta_1 k_x+\beta_3 {k_x}^3+\beta_4 k_x {k_y}^2}{2(\beta_2 k_y+\beta_6 {k_y}^3)}, \langle s_y\rangle = \frac{<\psi_2\|s_z\|\psi_2>}{<\psi_2\|\psi_2>} \approx \frac{1}{2}$ | | | | | | |
| | | $\beta_1$ | $\beta_2$ | $\beta_3$ | $\beta_4$ | $\beta_5$ | $\beta_6$ |
| | $k_y > k_{y-c}$ | -0.0000843367 | 0.00305 | -0.073645 | 0.65979 | 515.1 | 43.4076 |
| | $k_y < k_{y-c}$ | 0.000236667 | -0.08065 | -0.04521 | 3.20833 | -487.95 | -49.777 |
| $AgClO_4$ | $H = \beta_1 k_x\sigma_x + \beta_2 k_y\sigma_y + \beta_3 {k_x}^3\sigma_x + \beta_4 k_x {k_y}^2\sigma_x + \beta_5 k_y {k_x}^2\sigma_y + \beta_6 {k_y}^3\sigma_y$ <br> $\langle s_x\rangle = \frac{<\psi_1\|s_y\|\psi_1>}{<\psi_1\|\psi_1>} \approx -\frac{1}{2}, \langle s_y\rangle = \frac{<\psi_1\|s_z\|\psi_1>}{<\psi_1\|\psi_1>} \approx -\frac{\beta_2 k_y+\beta_5 k_y {k_x}^2+\beta_6 {k_y}^3}{2(\beta_1 k_x+\beta_3 {k_x}^3)}$ | | | | | | |
| | | $\beta_1$ | $\beta_2$ | $\beta_3$ | $\beta_4$ | $\beta_5$ | $\beta_6$ |
| | $k_x > k_{x-v}$ | 0.07608 | -0.000120075 | 63.54167 | 132.5 | 0.8188 | 0.023465 |
| | $k_x < k_{x-v}$ | -0.02558 | -0.000828753 | -63.54167 | -137.65 | 4.06172 | 0.00264 |

**TABLE S4.** Types of spin textures governed by linear-in-$k$ SOC Hamiltonians in 21 noncentrosymmetric point groups.[14] 13 point groups that support the high-quality PSTs predicted by our models are highlighted in bold. The intersection between crystal point group symmetry (CPGS) and wavevector point group symmetry (WPGS) is discriminated by different colors. Pink, blue and green means that the spin texture around the $k$ point with WPGS is governed by the typical Rashba, Weyl or Dresselhaus spin-orbit field, respectively, which are also labelled in the matrix element. Deep orange indicates that the spin texture is dictated by the inhomogeneous spin-orbit field (i.e., the coexistence of typical Rashba and Dresselhaus (RD) or Weyl and Dresselhaus (WD) spin-orbit fields). For some $k$ points with $C_n$ and $C_s$, WPGS where the $n$-fold rotation axis is located in the $k$ plane or the mirror is perpendicular to the $k$ plane, the in-plane spin texture is also dominated by the inhomogeneous spin-orbit field while the out-of-plane spin component will appear, which are labelled with shallow orange. U represents the undefined spin textures and CPST represents PSTs governed by $k$-cubic terms.

| | | | Crystal point group symmetry | | | | | | | | | | | | | | | | | | | | |
|---|---|---|---|---|---|---|---|---|---|---|---|---|---|---|---|---|---|---|---|---|---|---|---|
| | | | Polar Chiral | | | | | Polar Non-chiral | | | | | Non-polar Chiral | | | | | | Non-polar Non-chiral | | | | |
| | | | $C_1$ | $C_2$ | $C_3$ | $C_4$ | $C_6$ | $C_s$ | $\mathbf{C_{2v}}$ | $\mathbf{C_{3v}}$ | $\mathbf{C_{4v}}$ | $\mathbf{C_{6v}}$ | $\mathbf{D_2}$ | $\mathbf{D_3}$ | $\mathbf{D_4}$ | $\mathbf{D_6}$ | $\mathbf{T}$ | $\mathbf{O}$ | $S_4$ | $C_{3h}$ | $\mathbf{D_{2d}}$ | $\mathbf{D_{3h}}$ | $\mathbf{T_d}$ |
| Wavevector point group symmetry | Polar Chiral | $C_1$ | | | | | | | | | | | | | | | | | | | | | |
| | | $C_2$ | | | | | | | | | | | | | | | | | | | | | |
| | | $C_3$ | | | | | | | | | | | | | | | | | | | | | |
| | | $C_4$ | | | | | | | | | | | | | | | | | | | | | |
| | | $C_6$ | | | | | | | | | | | | | | | | | | | | | |
| | Polar Non-chiral | $C_s$ | | | | | | | | | | | | | | | | | | | | | |
| | | $C_{2v}$ | | | | | | | *RD* | | *RD* | *RD* | | | | | | | | | *RD* | *RD* | *RD* |
| | | $C_{3v}$ | | | | | | | | *R* | | *R* | | | | | | | | | | *R* | *R* |
| | | $C_{4v}$ | | | | | | | | | *R* | | | | | | | | | | | | |
| | | $C_{6v}$ | | | | | | | | | | *R* | | | | | | | | | | | |
| | Non-polar Chiral | $D_2$ | | | | | | | | | | | *WD* | | *WD* | *WD* | *WD* | *WD* | | | *WD* | | |
| | | $D_3$ | | | | | | | | | | | | *W* | | *W* | | *W* | | | | | |
| | | $D_4$ | | | | | | | | | | | | | *W* | | | *W* | | | | | |
| | | $D_6$ | | | | | | | | | | | | | | *W* | | | | | | | |
| | | $T$ | | | | | | | | | | | | | | | *W* | *W* | | | | | |
| | | $O$ | | | | | | | | | | | | | | | | *W* | | | | | |
| | Non-polar Non-chiral | $S_4$ | | | | | | | | | | | | | | | | | *U* | | *U* | | *U* |
| | | $C_{3h}$ | | | | | | | | | | | | | | | | | | *CPST* | | *CPST* | |
| | | $D_{2d}$ | | | | | | | | | | | | | | | | | | | *D* | | *D* |
| | | $D_{3h}$ | | | | | | | | | | | | | | | | | | | | *CPST* | |
| | | $T_d$ | | | | | | | | | | | | | | | | | | | | | *U* |

**TABLE S5.** All the *k* paths that possibly exhibit high-quality PSTs in different space groups. Here all the *k* points and coordinates are labelled according to the "The k-vector types and Brillouin zones of the space groups" of Bilbao Crystallographic Server.[15-17] RR/DD means the PSTs along the *k* path may be formed by two (inhomogeneous) Rashba or Dresselhaus spin-orbit fields, which is the same for WW/DD. The *k* points and lines satisfying our models are all listed, where the spin textures are dictated by either typical or inhomogeneous spin-orbit fields. Some *k* paths emphasized with bold font, indicating that within the *k* planes spanning them, the spin textures have no out-of-plane component. The *k* paths that have no symmetry are underlined with lines.

| Point Groups | Space groups | | Coordinates of *k* points (conventional basis) | *k* lines (typical) | *k* lines (inhomogeneous) |
|---|---|---|---|---|---|
| $C_{2v}$ | 25 | *Pmm*2 | ***RD*($C_{2v}$):**<br>Γ(0,0,0), Z(0,0,1/2),<br>Y(0,1/2,0), T(0,1/2,1/2), X(1/2,0,0),<br>U(1/2,0,1/2)<br>S(1/2,1/2,0), R(1/2,1/2,1/2) | | ***RR/DD:***<br>**Γ-X, Γ-Y,**<br>**Y-S, X-S,**<br>Z-T**, T-R,**<br>**U-R** |
| | 26 | $Pmc2_1$ | | | |
| | 27 | *Pcc*2 | | | |
| | 28 | *Pma*2 | | | |
| | 29 | $Pca2_1$ | | | |
| | 30 | *Pnc*2 | | | |
| | 31 | $Pmn2_1$ | | | |
| | 32 | *Pba*2 | | | |
| | 33 | $Pna2_1$ | | | |
| | 34 | *Pnn*2 | | | |
| | 35 | *Cmm*2 | a < b:<br>***RD*($C_{2v}$):**<br>Γ(0,0,0), Z(0,0,1/2),<br>Y(0,1,0), T(0,1,1/2)<br>***RD*($C_s$):**<br>SM(2u,0,0), A(2u,0,1/2), N(2u,1,0),<br>N2(2u,1,1/2)<br>***WD*($C_2$):**<br>S(1/2,1/2,0), R(1/2,1/2,1/2) | | ***RR/DD:***<br>**Γ-SM**, **Γ-Y**,<br>A-Z, **Y-N1**,<br>N1-N2, Z-T<br><br>***WW/DD:***<br>S-R |
| | 36 | $Cmc2_1$ | | | |
| | 37 | *Ccc*2 | | | |
| | 38 | *Amm*2 | a < b:<br>***RD*($C_{2v}$):**<br>Γ(0,0,0), Y(0,1,0),<br>T(1/2,1,0), Z(1/2,0,0),<br>SM(0,0,-2u), A(1/2,0,-2u)<br>***RD*($C_s$):**<br>S(0,1/2,-1/2), R(1/2,1/2,-1/2),<br>N1(0,1,-2u), N2(1/2,1,-2u) | | ***RR/DD:***<br>**S-R, Γ-Z,**<br>**Γ-Y, N1-N2,**<br>**Y-T, Z-T** |
| | 39 | *Aem*2 | | | |
| | 40 | *Ama*2 | | | |
| | 41 | *Aea*2 | | | |
| | 42 | *Fmm*2 | a>b+c:<br>***RD*($C_{2v}$):**<br>Γ(0,0,0), Z(0,0,1),<br>Y(0,1,0), T(0,1,1)<br>***RD*($C_s$):**<br>SM(2u,0,0), C(2u,1,0), A(2u,0,1) | | ***RR/DD:***<br>**Γ-Y, Z-T.**<br>**Γ-SM, SM-C,**<br>**C-Y, A-Z** |
| | 43 | *Fdd*2 | | | |

<table>
<tr><td rowspan="3"></td><td>44</td><td>Imm2</td><td rowspan="3">c>b>a or c>a>b:<br>***RD***($C_{2v}$):<br>Γ(0,0,0), X(0,0,1),<br>SM(2u,0,0)<br>***RD***($C_s$):<br>R(1/2,0,1/2), F(2u,0,1), DT(0,2u,0),<br>S(0,1/2,1/2)<br>***WD***($C_2$):<br>T(1/2,1/2,0), W(1/2,1/2,1/2)</td><td rowspan="3"></td><td rowspan="3">***RR/DD:***<br>**Γ-SM, F-X,**<br>**Γ-X, Γ-DT**<br>***WW/DD:***<br>W-T<br>***DD:***<br>W-R, S-W</td></tr>
<tr><td>45</td><td>Iba2</td></tr>
<tr><td>46</td><td>Ima2</td></tr>
<tr><td rowspan="6">$C_{3v}$</td><td>156</td><td>P3m1</td><td>***R***($C_{3v}$):<br>Γ(0,0,0), A(0,0,1/2)<br>***RD***($C_s$):<br>M(1/2,0,0), L(1/2,0,1/2)<br>***WD***($C_3$):<br>H(1/3,1/3,1/2), K(1/3,1/3,0)</td><td></td><td>***RR:***<br>Γ-M, A-L<br>***RR/DD:***<br>L-M<br>***WW/DD:***<br>H-K</td></tr>
<tr><td>157</td><td>P31m</td><td>***R***($C_{3v}$):<br>Γ(0,0,0), A(0,0,1/2),<br>K(1/3,1/3,0), H(1/3,1/3,1/2)<br>***RD***($C_s$):<br>M(1/2,0,0), L(1/2,0,1/2)</td><td>***RR:***<br>Γ-K, A-H</td><td>***RR:***<br>Γ-M, M-K,<br>L-H, A-L<br>***RR/DD:***<br>L-M</td></tr>
<tr><td>158</td><td>P3c1</td><td>***R***($C_{3v}$):<br>Γ(0,0,0), A(0,0,1/2)<br>***RD***($C_s$):<br>M(1/2,0,0), L(1/2,0,1/2)<br>***WD***($C_3$):<br>H(1/3,1/3,1/2), K(1/3,1/3,0)</td><td></td><td>***RR:***<br>Γ-M, A-L,<br>***RR/DD:***<br>L-M,<br>***WW/DD:***<br>H-K</td></tr>
<tr><td>159</td><td>P31c</td><td>***R***($C_{3v}$):<br>Γ(0,0,0), A(0,0,1/2),<br>K(1/3,1/3,0), H(1/3,1/3,1/2)<br>***RD***($C_s$):<br>M(1/2,0,0), L(1/2,0,1/2)</td><td>***RR:***<br>Γ-K, A-H</td><td>***RR:***<br>Γ-M, M-K,<br>L-H, A-L<br>***RR/DD:***<br>L-M</td></tr>
<tr><td>160</td><td>R3m</td><td rowspan="2">sqrt3a < sqrt2c:<br>***R***($C_{3v}$): Γ(0,0,0), T(0,1,1/2)<br>***RD***($C_s$):<br>L(-1/2,1/2,1/2), FB(0,1/2,1), CA(0,-u+v,-2u-v), C(u-v,-u+v, 2u+v)</td><td rowspan="2"></td><td rowspan="2">***RR/DD:***<br>FB-CA</td></tr>
<tr><td>161</td><td>R3c</td></tr>
<tr><td rowspan="9">$C_{4v}$</td><td>99</td><td>P4mm</td><td rowspan="8">***R***($C_{4v}$):<br>Γ(0,0,0), Z(0,0,1/2),<br>M(1/2,1/2,0), A(1/2,1/2,1/2)<br>***RD***($C_{2v}$):<br>X(0,1/2,0), R(0,1/2,1/2)</td><td rowspan="8">***RR:***<br>**Γ-M, Z-A**</td><td rowspan="8">***RR:***<br>**Γ-X, Z-R,**<br>**X-M, A-R**</td></tr>
<tr><td>100</td><td>P4bm</td></tr>
<tr><td>101</td><td>P42cm</td></tr>
<tr><td>102</td><td>P42nm</td></tr>
<tr><td>103</td><td>P4cc</td></tr>
<tr><td>104</td><td>P4nc</td></tr>
<tr><td>105</td><td>P42mc</td></tr>
<tr><td>106</td><td>P42bc</td></tr>
<tr><td>107</td><td>I4mm</td><td>c/a > 1:<br>***R***($C_{4v}$):</td><td></td><td>***RR:***<br>**Γ-X, Γ-SM,**</td></tr>
</table>

| | | | | | |
|---|---|---|---|---|---|
| | 108 | $I4cm$ | Γ(0,0,0), M(0,0,1)<br>***RD*(*$C_{2v}$*)**:<br>X(1/2,1/2,0), P(1/2,1/2,1/2)<br>***RD*(*$C_s$*)**:<br>Y(1/2+u,1/2-u,0), SM(2u,0,0),<br>F(2u,0,1), N(1/2,0,1/2),<br>U(u,u,1) | | **M-F, U-M**<br>***RR/DD:***<br>**X-Y, Y-SM,**<br>**N-P,** |
| | 109 | $I4_1md$ | | | |
| | 110 | $I4_1cd$ | | | |
| $C_{6v}$ | 183 | $P6mm$ | ***R*(*$C_{4v}$*)**:<br>Γ(0,0,0), A(0,0,1/2)<br>***R*(*$C_{3v}$*)**:<br>K(1/3,1/3,0), H(1/3,1/3,1/2)<br>***RD*(*$C_{2v}$*)**:<br>M(1/2,0,0), L(1/2,0,1/2) | ***RR:***<br>**Γ-K, A-H** | ***RR:***<br>**Γ-M, A-L,**<br>**K-M, L-H** |
| | 184 | $P6cc$ | | | |
| | 185 | $P6_3cm$ | | | |
| | 186 | $P6_3mc$ | | | |
| $D_2$ | 16 | $P222$ | ***WD*(*$D_2$*)**:<br>Γ(0,0,0), X(1/2,0,0),<br>Y(0,1/2,0), Z(0,0,1/2),<br>S(1/2,1/2,0), U(1/2,0,1/2),<br>T(0,1/2,1/2), R(1/2,1/2,1/2) | | ***WW/DD:***<br>**Γ-X, X-S,**<br>**S-Y, Y-G,**<br>**G-Z,** Z-U,<br>**U-R, R-T,**<br>T-Z, **Y-T,**<br>**U-X, S-R** |
| | 17 | $P222_1$ | | | |
| | 18 | $P2_12_12$ | | | |
| | 19 | $P2_12_12_1$ | | | |
| | 20 | $C222_1$ | a < b:<br>***WD*(*$D_2$*)**:<br>Γ(0,0,0), Y(0,1,0),<br>T(0,1,1/2), Z(0,0,1/2)<br>***WD*(*$C_2$*)**:<br>SM(2u,0,0), S(1/2,1/2,0),<br>R(1/2,1/2,1/2), A(2u,0,1/2),<br>C(2u',1,0), E(2u',1,1/2) | | ***WW/DD:***<br>**Γ-SM, S-R,**<br>A-Z, **Γ-Z,**<br>**Γ-Y, Y-C,**<br>**C-E, E-T,**<br>**T-Y,** Z-T |
| | 21 | $C222$ | | | |
| | 22 | $F222$ | $a^{-2} < b^{-2}+c^{-2}$, $b^{-2} < c^{-2}+a^{-2}$ and $c^{-2} < a^{-2}+b^{-2}$:<br>***WD*(*$D_2$*)**:<br>others:<br>Γ(0,0,0), Y(0,1,0),<br>T(1,0,0), Z(0,0,1)<br>***WD*(*$C_2$*)**:<br>C(2u,1,0), D(1,2u,0),<br>B(0,2u,1), H(0,1,2u),<br>A(2u,0,1), G(1,0,2u) | | ***WW/DD:***<br>**Γ-Y, Y-C,**<br>**C-D, D-T,**<br>**Γ-T, Γ-Z,**<br>**Z-B, B-H,**<br>**H-C, A-Z,**<br>**G-T, H-Y** |
| | 23 | $I222$ | c>b>a or c>a>b:<br>***WD*(*$D_2$*)**:<br>Γ(0,0,0), X(0,0,1),<br>W(1/2,1/2,1/2) | | ***WW/DD:***<br>**Γ-SM, W-T,**<br>**W-R, F-X,**<br>**Γ-X, Γ-DT,** |
| | 24 | $I2_12_12_1$ | **$WD(C_2)$:**<br>SM(2u,0,0), T(1/2,1/2,0),<br>R(1/2,0,1/2), F(2u,0,1),<br>DT(0,2u,0), S(0,1/2,1/2),<br>U(0,2u,1) | | **S-W, U-X** |
| $D_3$ | 149 | $P312$ | **$W(D_3)$:**<br>Γ(0,0,0), A(0,0,1/2)<br>**$WD(C_3)$:**<br>K(1/3,1/3,0), H(1/3,1/3,1/2)<br>**$WD(C_2)$:**<br>M(1/2,0,0), L(1/2,0,1/2) | | ***WW/DD:***<br>Γ-A, Γ-M,<br>K-H<br>***WW:***<br>A-L |
| | 150 | $P321$ | **$W(D_3)$:**<br>Γ(0,0,0), A(0,0,1/2),<br>K(1/3,1/3,0), H(1/3,1/3,1/2)<br>**$WD(C_2)$:**<br>M(1/2,0,0), L(1/2,0,1/2) | ***WW:***<br>Γ-K, A-H | ***WW/DD:***<br>**Γ-A**, **H-K**,<br>M-K, A-L<br>***WW:***<br>Γ-M, L-H |
| | 151 | $P3_112$ | **$W(D_3)$:**<br>Γ(0,0,0), A(0,0,1/2)<br>**$WD(C_3)$:**<br>K(1/3,1/3,0), H(1/3,1/3,1/2)<br>**$WD(C_2)$:**<br>M(1/2,0,0), L(1/2,0,1/2) | | ***WW/DD:***<br>Γ-A, Γ-M,<br>K-H<br>***WW:***<br>A-L |
| | 152 | $P3_121$ | **$W(D_3)$:**<br>Γ(0,0,0), A(0,0,1/2),<br>K(1/3,1/3,0), H(1/3,1/3,1/2)<br>**$WD(C_2)$:** M(1/2,0,0), L(1/2,0,1/2) | ***WW:***<br>Γ-K, A-H | ***WW/DD:***<br>**Γ-A**, **H-K**,<br>M-K, A-L<br>***WW:***Γ-M, L-H |
| | 153 | $P3_212$ | **$W(D_3)$:**<br>Γ(0,0,0), A(0,0,1/2)<br>**$WD(C_3)$:**<br>K(1/3,1/3,0), H(1/3,1/3,1/2)<br>**$WD(C_2)$:**<br>M(1/2,0,0), L(1/2,0,1/2) | | ***WW/DD:***<br>Γ-A, Γ-M,<br>K-H<br>***WW:***<br>A-L |
| | 154 | $P3_221$ | **$W(D_3)$:**<br>Γ(0,0,0), A(0,0,1/2),<br>K(1/3,1/3,0), H(1/3,1/3,1/2)<br>**$WD(C_2)$:**<br>M(1/2,0,0), L(1/2,0,1/2) | ***WW:***<br>Γ-K, A-H | ***WW/DD:***<br>**Γ-A**, **H-K**,<br>M-K, A-L<br>***WW:***<br>Γ-M, L-H |
| | 155 | $R32$ | sqrt3a < sqrt2c:<br>**$W(D_3)$:**<br>Γ(0,0,0), T(0,0,3/2)<br>**$WD(C_2)$:**<br>YA(1/3+u,1/3-2u,1/2), SN(u,u,0),<br>QA(2u,1/2-u,1), FB(0,1/2,1) | | ***WW/DD:***<br>**Γ-T,** Γ-SN<br>SN-QA |
| $D_4$ | 89 | $P422$ | **$W(D_4)$:**<br>Γ(0,0,0), Z(0,0,1/2),<br>M(1/2,1/2,0), A(1/2,1/2,1/2) | ***WW:***<br>**Γ-M,** Z-A | ***WW/DD:***<br>**Γ-Z**, **M-A**,<br>**Γ-X**, Z-R, |
| | 90 | $P42_12$ | | | |
| | 91 | $P4_122$ | | | |
| | 92 | $P4_12_12$ | | | |
| | 93 | $P4_222$ | | | |
| | 94 | $P4_22_12$ | ***WD*($D_2$):**<br>X(0,1/2,0), R(0,1/2,1/2) | | **M-X. A-R**<br>**X-R** |
| | 95 | $P4_322$ | | | |
| | 96 | $P4_32_12$ | | | |
| | 97 | $I422$ | c/a < 1:<br>***W*($D_4$):**<br>Γ(0,0,0), M(0,0,1)<br>***WD*($D_2$):**<br>X(1/2,1/2,0), P(1/2,1/2,1/2)<br>***WD*($C_4$):**<br>LD(0,0,2u), V(1,0,2u)<br>***WD*($C_2$):**<br>N(1/2,0,1/2) | | ***WW/DD:***<br>**Γ-M, Γ-LD,**<br>**P-N, N-V,**<br>V-M**, X-P**<br>***WW*：**<br>**Γ-X** |
| | 98 | $I4_122$ | | | |
| $D_6$ | 177 | $P622$ | ***W*($D_6$):**<br>Γ(0,0,0), A(0,0,1/2)<br>***W*($D_3$):**<br>K(1/3,1/3,0), H(1/3,1/3,1/2)<br>***WD*($D_2$):**<br>M(1/2,0,0), L(1/2,0,1/2) | ***WW:***<br>**Γ-K**, **H-A** | ***WW/DD:***<br>**Γ-A**, **K-H**,<br>**M-L**<br>***WW:***<br>**Γ-M, A-L,**<br>**K-M, H-L** |
| | 178 | $P6_122$ | | | |
| | 179 | $P6_522$ | | | |
| | 180 | $P6_222$ | | | |
| | 181 | $P6_422$ | | | |
| | 182 | $P6_322$ | | | |
| $T$ | 195 | $P23$ | ***W(T)*:**<br>Γ(0,0,0), R(1/2,1/2,1/2)<br>***WD*($D_2$):**<br>M(1/2,1/2,0), X(0,1/2,0) | ***WW:***<br>**Γ-R** | ***WW:***<br>**Γ-X, R-M,**<br>**Γ-M, R-X**<br>***WW/DD*：**<br>**M-X** |
| | 196 | $F23$ | ***W(T)*:**<br>Γ(0,0,0)<br>***WD*($D_2$):**<br>X(0,1,0)<br>***WD*($C_3$):**<br>L(1/2,1/2,1/2)<br>***WD*($C_2$):**<br>W(1/2,1,0) | | ***WW:***<br>**Γ-X,** Γ-L<br>***WW/DD*：**<br>**X-W** |
| | 197 | $I23$ | ***W(T)*:**<br>Γ(0,0,0), H(0,1,0),<br>P(1/2,1/2,1/2)<br>***WD*($C_2$):**<br>N(1/2,1/2,0) | ***WW:***<br>**Γ-H,** Γ-P | ***WW:***<br>**H-N, Γ-N,**<br>**P-N** |
| | 198 | $P2_13$ | ***W(T)*:**<br>Γ(0,0,0), R(1/2,1/2,1/2)<br>***WD*($D_2$):**<br>M(1/2,1/2,0), X(0,1/2,0) | ***WW:***<br>**Γ-R** | ***WW:***<br>**Γ-X, R-M,**<br>**Γ-M, R-X**<br>***WW/DD*：**<br>**M-X** |
| | 199 | $I2_13$ | ***W(T)*:**<br>Γ(0,0,0), H(0,1,0),<br>P(1/2,1/2,1/2)<br>***WD*($C_2$):**<br>N(1/2,1/2,0) | ***WW:***<br>**Γ-H,** Γ-P | ***WW:***<br>**H-N, Γ-N,**<br>**P-N** |
| $O$ | 207 | $P432$ | ***W(T)***:<br>Γ(0,0,0), R(1/2,1/2,1/2)<br>***W($D_4$)***:<br>M(1/2,1/2,0), X(0,1/2,0) | ***WW:***<br>**Γ-M, Γ-X,**<br>**X-M, Γ-R** | ***WW:***<br>**R-M, R-X** |
| | 208 | $P4_232$ | | | |
| | 209 | $F432$ | ***W(O)***: Γ(0,0,0)<br>***W($D_4$)***: X(0,1,0)<br>***W($D_3$)***: L(1/2,1/2,1/2)<br>***WD($D_2$)***: W(1/2,1,0)<br>***WD($C_2$)***: SM(2u,2u,0), S(2u,1,2u) | ***WW:***<br>**Γ-X,** Γ-L | ***WW:***<br>**X-W**, **Γ-SM**<br>***WW/DD:***<br>**W-SM, S-W,**<br>**W-L, S-X** |
| | 210 | $F4_132$ | | | |
| | 211 | $I432$ | ***W(O)***:<br>Γ(0,0,0), H(0,1,0)<br>***W(T)***:<br>P(1/2,1/2,1/2)<br>***WD($D_2$)***:<br>N(1/2,1/2,0) | ***WW:***<br>**Γ-H,** G-P**, P-H** | ***WW:***<br>**Γ-N, H-N,**<br>**P-N** |
| | 212 | $P4_332$ | ***W(T)***:<br>Γ(0,0,0), R(1/2,1/2,1/2)<br>***W($D_4$)***:<br>M(1/2,1/2,0), X(0,1/2,0) | ***WW:***<br>**Γ-M, Γ-X,**<br>**X-M, Γ-R** | ***WW:***<br>**R-M, R-X** |
| | 213 | $P4_132$ | | | |
| | 214 | $I4_132$ | ***W(O)***:<br>Γ(0,0,0), H(0,1,0)<br>***W(T)***:<br>P(1/2,1/2,1/2)<br>***WD($D_2$)***:<br>N(1/2,1/2,0) | ***WW:***<br>**Γ-H,** G-P**, P-H** | ***WW:***<br>**Γ-N, H-N,**<br>**P-N** |
| $D_{2d}$ | 111 | $P$-42$m$ | ***D($D_{2d}$)***:<br>Γ(0,0,0), A(1/2,1/2,1/2)<br>Z(0,0,1/2), M(1/2,1/2,0)<br>***WD($D_2$)***:<br>X(0,1/2,0), R(0,1/2,1/2) | ***DD:***<br>**Γ-M, A-Z** | ***DD:***<br>**Γ-X, A-R,**<br>**Z-R, X-M** |
| | 112 | $P$-42$c$ | | | |
| | 113 | $P$-42$_1m$ | | | |
| | 114 | $P$-42$_1c$ | | | |
| | 115 | $P$-4$m$2 | ***D($D_{2d}$)***:<br>Γ(0,0,0), A(1/2,1/2,1/2)<br>Z(0,0,1/2), M(1/2,1/2,0)<br>***RD($C_{2v}$)***:<br>X(0,1/2,0), R(0,1/2,1/2) | ***DD:***<br>**Γ-M, A-Z** | ***DD:***<br>**Γ-X, A-R,**<br>**Z-R, X-M** |
| | 116 | $P$-4$c$2 | | | |
| | 117 | $P$-4$b$2 | | | |
| | 118 | $P$-4$n$2 | | | |
| | 119 | $I$-4$m$2 | c/a > 1:<br>***D($D_{2d}$)***:<br>Γ(0,0,0), M(0,0,1)<br>***WD($D_2$)***:<br>X(0,1/2,0) | | ***DD:***<br>**Γ-X, Γ-SM,**<br>**M-F, U-M**<br>***RR/DD:***<br>**N-F** |
|  | 120 | *I*-4*c*2 | ***RD($C_s$):***<br>SM(2u,0,0), F(2u,0,1), N(1/2,0,1/2)<br>***WD($C_2$):***<br>Y(1/2+u,1/2-u,0), U(u,u,1) |  |  |
|  | 121 | *I*-42*m* | c/a > 1:<br>***D($D_{2d}$)*:**<br>Γ(0,0,0), M(0,0,1)<br>P(1/2,1/2,1/2)<br>***RD($C_{2v}$):***<br>X(1/2,1/2,0)<br>***RD($C_s$):***<br>Y(1/2+u,1/2-u,0), U(u,u,1)<br>***WD($C_2$):***<br>SM(2u,0,0), F(2u,0,1), N(1/2,0,1/2) |  | ***DD:***<br>**Γ-X, Y-SM,**<br>**Γ-SM, M-F,**<br>**N-P**, U-M<br>***RR/DD:***<br>**X-Y**<br>***WW/DD:***<br>**F-N** |
|  | 122 | *I*-42*d* |  |  |  |
| $D_{3h}$ | 187 | *P*-6*m*2 | ***RD($C_{2v}$)*:**<br>M(1/2,0,0), L(1/2,0,1/2) |  | ***RR/DD:***<br>**L-M** |
|  | 188 | *P*-6*c*2 |  |  |  |
|  | 189 | *P*-62*m* | ***RD($C_{2v}$)*:**<br>M(1/2,0,0), L(1/2,0,1/2) |  | ***RR/DD:***<br>**L-M** |
|  | 190 | *P*-62*c* |  |  |  |
| $T_d$ | 215 | *P*-43*m* | ***D($D_{2d}$)*:**<br>M(1/2,1/2,0), X(0,1/2,0) | ***DD:***<br>**X-M** |  |
|  | 216 | *F*-43*m* | ***D($D_{2d}$)*:**<br>X(0,1,0)<br>***R($C_{3v}$)*:**<br>L(1/2,1/2,1/2)<br>***RD($C_s$)*:**<br>S(2u,1,2u) |  | ***DD:*** **X-S** |
|  | 217 | *I*-43*m* | ***RD($C_{2v}$)*:**<br>N(1/2,1/2,0) |  |  |
|  | 218 | *P*-43*n* | ***D($D_{2d}$)*:**<br>M(1/2,1/2,0), X(0,1/2,0) | ***DD:***<br>**X-M** |  |
|  | 219 | *F*-43*c* | ***D($D_{2d}$)*:**<br>X(0,1,0)<br>***R($C_{3v}$)*:**<br>L(1/2,1/2,1/2)<br>***RD($C_s$)*:**<br>S(2u,1,2u) |  | ***DD:***<br>**X-S** |
|  | 220 | *I*-43*d* | ***RD($C_{2v}$)*:**<br>N(1/2,1/2,0) |  |  |

**TABLE S6.** The dynamically stable materials possessing high-quality PSTs with various point groups. Only a subset of materials is listed here, where the CBM/VBM lies at the midpoint of the required *k* paths. Note that for some materials, the spin-orbit fields around the *k* points may be inhomogeneous, which does not greatly affect the quality of the PST due to the generality of our models.

| Location of CBM/VBM | Materials | Space group | *k*-path | PST-Type | Band Gaps (eV) |
|---|---|---|---|---|---|
| **CBM/VBM located exactly in the PST region** | $RbBiNb_2O_7$ | $Pmc2_1$ ($C_{2v}$) | X-S | DD-2 | 2.36 |
| | $RbTlS_2O_8$ | $R32$ ($D_3$) | Γ-T | WW | 2.2 |
| | $CaTa_4O_{11}$ | $P6_322$ ($D_6$) | H-K | DD-1 | 3.36 |
| | $SrTa_4O_{11}$ | $P6_322$ ($D_6$) | H-K | DD-1 | 3.34 |
| | $Na_2Sn_2O_3$ | $I2_13$ ($T$) | Γ-H | WW | 0.08 |
| | $GeBi_{12}O_{20}$ | $I23$ ($T$) | Γ-H | WW | 2.04 |
| | $SiBi_{12}O_{20}$ | $I23$ ($T$) | Γ-H | WW | 2.04 |
| | $TiBi_{12}O_{20}$ | $I23$ ($T$) | Γ-H | WW | 1.97 |
| | GeRu | $P2_13$ ($T$) | Γ-X | WW | 0.06 |
| | $AgClO_4$ | $I$-42$m$ ($D_{2d}$) | Γ-Σ | DD-1 | 2.3 |
| **CBM/VBM close to the PST region** | AgH | $P6_3mc$ ($C_{6v}$) | Γ-K | RR | 0.71 |
| | $Hg_3AsS_4Cl$ | $P6_3mc$ ($C_{6v}$) | Γ-K | RR | 1.36 |
| | NaSnP | $P6_3mc$ ($C_{6v}$) | Γ-K | RR | 0.51 |
| | $K_2Sn_2O_3$ | $I2_13$ ($T$) | Γ-H | WW | 0.94 |
| | $K_2Pb_2O_3$ | $I2_13$ ($T$) | Γ-H | WW | 1.03 |
| | $Rb_2Pb_2O_3$ | $I2_13$ ($T$) | Γ-H | WW | 1.1 |
| | $Cs_2Pb_2O_3$ | $I2_13$ ($T$) | Γ-H | WW | 1.04 |
| | $Li_2GePbS_4$ | $I$-42$m$ ($D_{2d}$) | Γ-Σ | DD-1 | 1.51 |

**TABLE S7.** High symmetry *k* points used in the calculations of the band structure for various compounds, given in units of the reciprocal lattice vectors.

| $Na_2Sn_2O_3$ | |
|---|---|
| ***k* point label** | **wavevector** |
| **Γ** | 0 0 0 |
| **H** | 1/2 -1/2 1/2 |
| **N** | 0 0 1/2 |
| **P** | 1/4 1/4 1/4 |

| $AgClO_4$ | |
|---|---|
| ***k* point label** | **wavevector** |
| **Γ** | 0 0 0 |
| **X** | 0 0 1/2 |
| **Y** | -0.25691 0.25691 0.5 |
| **Σ** | -0.37845 0.37845 0.37845 |
| **Z** | 1/2 1/2 -1/2 |
| **Σ1** | 0.37845 0.62154 -0.37845 |
| **N** | 0 1/2 0 |
| **P** | 1/4 1/4 1/4 |
| **Y1** | 0.5 0.5 -0.25691 |

**TABLE S8.** $k{\cdot}p$ models and corresponding parameters for different $k$ points in $Na_2Sn_2O_3$. $\beta_1, \beta_2, \beta_3$ are the SOC coefficients for terms of $k_x$, $k_y, k_z$. $k_{x-c}/k_{y-c}$ represents the $k_x/k_y$ component of the location of CBM in the $k_x$-$k_y$ plane.

| $Na_2Sn_2O_3$ | | |
|---|---|---|
| **Cartesian coordinate** | $k{\cdot}p$ model ($k_x$-$k_y$ plane) | Parameter ($\beta_1, \beta_2, \beta_3$) (eV Å) |
| **Γ (0, 0, 0)** | $\alpha(k_x^2 + k_y^2) + \beta(\sigma_x k_x + \sigma_y k_y)$ | $(0.43085, 0.43085, 0)$ |
| **H (0, 1, 0)** | $\alpha(k_x^2 + k_y^2) + \beta(\sigma_x k_x + \sigma_y k_y)$ | $(-1.28016, -1.28016, 0)$ |
| **CBM$(0,\ k_{y-c}, 0)$** | $\alpha_1 k_x^2 + \alpha_2 k_y^2 + \beta_1 k_x \sigma_x + \beta_2 k_y \sigma_y$ | $(0.000192, 0.02201, 0)$ [$k_y > k_{y\text{-}c}$] $(0.00156, -0.1023, 0)$ [$k_y < k_{y\text{-}c}$] |

**TABLE S9.** $k{\cdot}p$ models and corresponding parameters for different $k$ points in $AgClO_4$ before and after applying pressure. $\beta_1, \beta_2, \beta_3$ are the SOC coefficients for terms of $k_x$,$k_y$, $k_z$. $k_{x-v}$ represents the $k_x$ component of the location of VBM in the $k_x$-$k_y$ plane.

| **$AgClO_4$** | | |
|---|---|---|
| **Cartesian coordinate** | $k{\cdot}p$ model ($k_x$-$k_y$ plane) | Parameter ($\beta_1, \beta_2, \beta_3$) (eV Å) |
| **Γ (0, 0, 0)** | $\alpha(k_x^2 + k_y^2) + \beta(\sigma_x k_x - \sigma_y k_y)$ | $(0.07249, 0.07249, 0)$ |
| **Σ (1, 0, 0)** | $\alpha_1 k_x^2 + \alpha_2 k_y^2 + \beta_1 k_x \sigma_x + \beta_2 k_y \sigma_y$ | $(-1.51584, 0.12393, 0)$ |
| VBM$(\boldsymbol{k_{x-v}}\ \boldsymbol{0}, \boldsymbol{0})$ | $\alpha_1 k_x^2 + \alpha_2 k_y^2 + \beta_1 k_x \sigma_x + \beta_2 k_y \sigma_y$ | $(0.10211, 0.000656, 0)$ [$k_x > k_{x-v}$] $(-0.05131, -0.000768, 0)$ [$k_x < k_{x-v}$] |

| **$AgClO_4$ (3GPa)** | | |
|---|---|---|
| **Cartesian coordinate** | $k{\cdot}p$ model ($k_x$-$k_y$ plane) | Parameter ($\beta_1, \beta_2, \beta_3$) (eV Å) |
| VBM$(\boldsymbol{k_{x-v}}\ \boldsymbol{0}, \boldsymbol{0})$ | $\alpha_1 k_x^2 + \alpha_2 k_y^2 + \beta_1 k_x \sigma_x + \beta_2 k_y \sigma_y$ | $(0.01172, 0.0000739, 0)$ [$k_x > k_{x-v}$] $(-0.14737, -0.000619, 0)$ [$k_x < k_{x-v}$] |

**TABLE S10.** Key metrics for various PST compounds. $k_{y-c}$ represents the $k_y$ component of the location of CBM in the $k_x$-$k_y$ plane. $k_{x-v}$ represents the $k_x$ component of the location of VBM in the $k_x$-$k_y$ plane.

| | **Materials** | | **△soc (meV)** | **Rashba anisotropy** | $\boldsymbol{S}(\text{Å}^{-2})$ | $\frac{\tau_s}{T_{\mathrm{PSH}}}$ | $\boldsymbol{l}_{\mathbf{PSH}}\ (\boldsymbol{nm})$ | **Band Gap (eV)** |
|---|---|---|---|---|---|---|---|---|
| **This work** | $Na_2Sn_2O_3$ | $k_y > k_{y\text{-}c}$ | 119 | 0.008 | **0.02** | **5253** | 77.8 | 0.08 |
| | | $k_y < k_{y\text{-}c}$ | | -0.015 | | **1513** | 16.7 | |
| | $AgClO_4$ | $k_x > k_{x\text{-}v}$ | 68 | 0.006 | **0.016** | **8387** | 23.5 | 2.3 |
| | | $k_x < k_{x\text{-}v}$ | | 0.015 | | **1577** | 46.7 | |
| | $AgClO_4$ (3 GPa) | $k_x > k_{x\text{-}v}$ | 78 | 0.006 | **0.0336** | **11655** | 214.0 | 2.31 |
| | | $k_x < k_{x\text{-}v}$ | | 0.004 | | **20200** | 17.0 | |
| **Previous Work[5]** | $BiInO_3$ | | 260 | 0.273 | 0.012 | 688 | 2.0 | 2.6 |
| | $Sr_3Zr_2O_7$ | | 22 | 0.105 | 0.037 | 238 | 9.2 | 3.7 |
| | $Sr_3Hf_2O_7$ | | 85 | 0.033 | 0.042 | 700000 | 2.5 | 4.1 |
| | $Ca_3Zr_2O_7$ | | 114 | 0.174 | 0.012 | 85 | 14.5 | 4.0 |
| | $Ca_3Hf_2O_7$ | | 156 | 0.068 | 0.025 | 620 | 6.7 | 4.4 |

**TABLE S11.** Parameters for computing the spin lifetime and PSH period in different compounds. Here $\alpha$ and $\beta$ are the SOC coefficients for the terms of $k_x$ and $k_y$, respectively.

| Materials | | | $\alpha$ (eVÅ) | $\beta$ (eVÅ) | $\tau_s$ (ps) | $T_{PSH}$ (ps) | $m^*/m_e$ | $k_F$ (Å$^{-1}$) |
|---|---|---|---|---|---|---|---|---|
| **This work** | $Na_2Sn_2O_3$ | $k_y > k_{y\text{-}c}$ | 0.000192 | 0.022 | **7407** | 1.41 | 1.4 | 0.067 |
| | | $k_y < k_{y\text{-}c}$ | 0.00156 | -0.102 | **454** | 0.3 | 1.4 | 0.067 |
| | $AgClO_4$ | $k_x > k_{x\text{-}v}$ | 0.102 | 0.000656 | **2516** | 0.3 | -1.0 | 0.067 |
| | | $k_x < k_{x\text{-}v}$ | -0.0513 | -0.000768 | **946** | 0.6 | -1.0 | 0.067 |
| | $AgClO_4$ (3 GPa) | $k_x > k_{x\text{-}v}$ | 0.0117 | 0.000073 | **30769** | 2.64 | -0.956 | 0.067 |
| | | $k_x < k_{x\text{-}v}$ | -0.147 | 9<br>-0.000619 | **4040** | 0.2 | -0.956 | 0.067 |
| **Previous Work[5]** | $BiInO_3$ | | 0.521 | 1.91 | 11 | 0.016 | 0.61 | 0.067 |
| | $Sr_3Zr_2O_7$ | | 0.0056 | 0.0534 | 138 | 0.58 | 4.9 | 0.067 |
| | $Sr_3Hf_2O_7$ | | 0.011 | 0.35 | 63000 | 0.089 | 2.69 | 0.067 |
| | $Ca_3Zr_2O_7$ | | 0.0122 | 0.07 | 37 | 0.44 | 2.4 | 0.067 |
| | $Ca_3Hf_2O_7$ | | 0.0145 | 0.214 | 93 | 0.15 | 1.68 | 0.067 |

**TABLE S12.** Comparison of distinctive types of PSTs.

| **Materials with PSTs** | **Quantum wells (GaAs/AlGaAs, InGaAs/InAlAs, etc.)** | **$Ge_3Pb_5O_{11}$, $BA_2PbCl_4$ monolayer, etc.** | **$Na_2Sn_2O_3$, $AgClO_4$, etc. (This work)** |
|---|---|---|---|
| **Mechanism of PST formation** | Balance of Rashba and Dresselhaus SOC effects | Protected by symmetry (governed by the single spin-orbit field) | **Interaction of two kinds spin-orbit fields** (in the midpoint of high-symmetry lines) |
| **Advantages of PST** | Small spin deviation (mainly from the terms cubic in *k*) | No need to regulate the strength of SOC effects | **1) Allow the appearance of high-quality PST** (small spin deviation, large area and long spin lifetime) **2) No need to regulate the strength of SOC effects** (intrinsic properties) **3) Abundant materials can be found efficiently in various point groups** (according to our design principle) |

**TABLE S13.** Irreducible representations of the Double Point Group *T*. The data can be found from the Bilbao Crystallographic Server.[15-17]

| | e | $2_{001}$<br>$2_{010}$<br>$2_{100}$<br>$\overline{2_{001}}$<br>$\overline{2_{010}}$<br>$\overline{2_{100}}$ | $3^+_{111}$<br>$3^+_{\bar{1}1\bar{1}}$<br>$3^+_{1\bar{1}\bar{1}}$<br>$3^+_{\bar{1}\bar{1}1}$ | $3^-_{111}$<br>$3^-_{\bar{1}1\bar{1}}$<br>$3^-_{1\bar{1}\bar{1}}$<br>$3^-_{\bar{1}\bar{1}1}$ | $\bar{e}$ | $\overline{3^+_{111}}$<br>$\overline{3^+_{\bar{1}1\bar{1}}}$<br>$\overline{3^+_{1\bar{1}\bar{1}}}$<br>$\overline{3^+_{\bar{1}\bar{1}1}}$ | $\overline{3^-_{111}}$<br>$\overline{3^-_{\bar{1}1\bar{1}}}$<br>$\overline{3^-_{1\bar{1}\bar{1}}}$<br>$\overline{3^-_{\bar{1}\bar{1}1}}$ |
|---|---|---|---|---|---|---|---|
| $A$ | 1 | 1 | 1 | 1 | 1 | 1 | 1 |
| ${}^1E$ | 1 | 1 | $-(1-i\sqrt{3})/2$ | $-(1+i\sqrt{3})/2$ | 1 | $-(1-i\sqrt{3})/2$ | $-(1+i\sqrt{3})/2$ |
| ${}^2E$ | 1 | 1 | $-(1+i\sqrt{3})/2$ | $-(1-i\sqrt{3})/2$ | 1 | $-(1+i\sqrt{3})/2$ | $-(1-i\sqrt{3})/2$ |
| $T$ | 3 | -1 | 0 | 0 | 3 | 0 | 0 |
| $\overline{E}$ | 2 | 0 | 1 | 1 | -2 | -1 | -1 |
| ${}^2\overline{F}$ | 2 | 0 | $-(1-i\sqrt{3})/2$ | $-(1+i\sqrt{3})/2$ | -2 | $(1-i\sqrt{3})/2$ | $(1+i\sqrt{3})/2$ |
| ${}^1\overline{F}$ | 2 | 0 | $-(1+i\sqrt{3})/2$ | $-(1-i\sqrt{3})/2$ | -2 | $(1+i\sqrt{3})/2$ | $(1-i\sqrt{3})/2$ |

**TABLE S14.** Characters of the representations of $D^{\frac{1}{2}}$.

| T | e | $2_{001}$<br>$2_{010}$<br>$2_{100}$<br>$\overline{2_{001}}$<br>$\overline{2_{010}}$<br>$\overline{2_{100}}$ | $3^+_{111}$<br>$3^+_{\bar{1}1\bar{1}}$<br>$3^+_{1\bar{1}\bar{1}}$<br>$3^+_{\bar{1}\bar{1}1}$ | $3^-_{111}$<br>$3^-_{\bar{1}1\bar{1}}$<br>$3^-_{1\bar{1}\bar{1}}$<br>$3^-_{\bar{1}\bar{1}1}$ | $\bar{e}$ | $\overline{3^+_{111}}$<br>$\overline{3^+_{\bar{1}1\bar{1}}}$<br>$\overline{3^+_{1\bar{1}\bar{1}}}$<br>$\overline{3^+_{\bar{1}\bar{1}1}}$ | $\overline{3^-_{111}}$<br>$\overline{3^-_{\bar{1}1\bar{1}}}$<br>$\overline{3^-_{1\bar{1}\bar{1}}}$<br>$\overline{3^-_{\bar{1}\bar{1}1}}$ |
|---|---|---|---|---|---|---|---|
| $D^{1/2}$ | 2 | 0 | 1 | 1 | -2 | -1 | -1 |

**TABLE S15.** The symmetry-adapted *k* terms according to the symmetries of *A* and *T*, which are obtained by utilizing the projection operator technique.

| | First-order | Second-order |
|---|---|---|
| A | \ | $k_x^2+k_y^2+k_z^2$ |
| T | $(k_z, k_x, k_y)$ | $(k_xk_y, k_yk_z, k_xk_z)$ |

**TABLE S16.** The C-G coefficients between the *irreps* $\bar{E}^1$and $\bar{E}^2$ through *irreps* $A$ and $T$.

| | A$\bar{E}^1$ | A$\bar{E}^2$ |
|---|---|---|
| $\bar{E}^1$ | 1 | 0 |
| $\bar{E}^2$ | 0 | 1 |

| | T$^1\bar{E}^1$ | T$^1\bar{E}^2$ |
|---|---|---|
| $\bar{E}^1$ | $1/\sqrt{3}$ | 0 |
| $\bar{E}^2$ | 0 | $-1/\sqrt{3}$ |

| | T$^2\bar{E}^1$ | T$^2\bar{E}^2$ |
|---|---|---|
| $\bar{E}^1$ | 0 | $1/\sqrt{3}$ |
| $\bar{E}^2$ | $1/\sqrt{3}$ | 0 |

| | T$^3\bar{E}^1$ | T$^3\bar{E}^2$ |
|---|---|---|
| $\bar{E}^1$ | 0 | $i/\sqrt{3}$ |
| $\bar{E}^2$ | $-i/\sqrt{3}$ | 0 |

**TABLE S17.** Irreducible representations of the Double Point Group $D_{2d}$.

| | e | $2_{001}$<br>$\overline{2_{001}}$ | $-4^{+}_{001}$<br>$-4^{-}_{001}$ | $2_{010}$<br>$2_{100}$<br>$\overline{2_{010}}$<br>$\overline{2_{100}}$ | $m_{110}$<br>$m_{1\bar{1}0}$<br>$\overline{m_{110}}$<br>$\overline{m_{1\bar{1}0}}$ | $\bar{e}$ | $\overline{-4^{+}_{001}}$<br>$\overline{-4^{-}_{001}}$ |
|---|---|---|---|---|---|---|---|
| $A_1$ | 1 | 1 | 1 | 1 | 1 | 1 | 1 |
| $B_1$ | 1 | 1 | -1 | 1 | -1 | 1 | -1 |
| $B_2$ | 1 | 1 | -1 | -1 | 1 | 1 | -1 |
| $A_2$ | 1 | 1 | 1 | -1 | -1 | 1 | 1 |
| $E$ | 2 | -2 | 0 | 0 | 0 | 2 | 0 |
| $\bar{E}_2$ | 2 | 0 | $\sqrt{2}$ | 0 | 0 | -2 | $\sqrt{2}$ |
| $\bar{E}_1$ | 2 | 0 | $\sqrt{2}$ | 0 | 0 | -2 | $\sqrt{2}$ |

**TABLE S18.** Characters of the representations of $D^{\frac{1}{2}}$.

| $D_{2d}$ | e | $2_{001}$<br>$\overline{2_{001}}$ | $-4^{+}_{001}$<br>$-4^{-}_{001}$ | $2_{010}$<br>$2_{100}$<br>$\overline{2_{010}}$<br>$\overline{2_{100}}$ | $m_{110}$<br>$m_{1\bar{1}0}$<br>$\overline{m_{110}}$<br>$\overline{m_{1\bar{1}0}}$ | $\bar{e}$ | $\overline{-4^{+}_{001}}$<br>$\overline{-4^{-}_{001}}$ |
|---|---|---|---|---|---|---|---|
| $D^{1/2}$ | 2 | 0 | $\sqrt{2}$ | 0 | 0 | -2 | -$\sqrt{2}$ |

**TABLE S19.** The symmetry-adapted *k* terms according to the symmetries of *A1*, *A2* and *E*, which are obtained by utilizing the projection operator technique.

| | First-order | Second-order |
|---|---|---|
| A1 | \ | $k_x^2 + k_y^2, k_z^2$ |
| A2 | \ | \ |
| E | $(k_x, k_y)$ | $(k_x k_z, k_y k_z)$ |

**TABLE S20.** The C-G coefficients between the *irreps* $\bar{E}^1$ and $\bar{E}^2$ through *irreps* $A_1$ and *E*.

| | $A_1\bar{E}^1$ | $A_1\bar{E}^2$ |
|---|---|---|
| $\bar{E}^1$ | 1 | 0 |
| $\bar{E}^2$ | 0 | 1 |

| | $E^2\bar{E}^1$ | $E^2\bar{E}^2$ |
|---|---|---|
| $\bar{E}^1$ | 0 | -1/$\sqrt{2}$ |
| $\bar{E}^2$ | -1/$\sqrt{2}$ | 0 |

| | $E^1\bar{E}^1$ | $E^1\bar{E}^2$ |
|---|---|---|
| $\bar{E}^1$ | 0 | i/$\sqrt{2}$ |
| $\bar{E}^2$ | -i/$\sqrt{2}$ | 0 |

**TABLE S21.** Irreducible representations of the Double Point Group $C_{2y}$.

| | e | $2_{010}$ | $\bar{e}$ | $\overline{2_{010}}$ |
|---|---|---|---|---|
| *A* | 1 | 1 | 1 | 1 |
| *B* | 1 | -1 | 1 | -1 |
| ${}^2\bar{E}$ | 1 | -i | -1 | i |
| ${}^1\bar{E}$ | 1 | i | -1 | -i |

**TABLE S22.** Characters of the representations of $D^{\frac{1}{2}}$.

| $C_{2y}$ | e | $2_{010}$ | $\bar{e}$ | $\overline{2_{010}}$ |
|---|---|---|---|---|
| $D^{1/2}$ | 2 | 0 | -2 | 0 |

**TABLE S23.** The symmetry-adapted $k$ terms according to the symmetries of $A$ and $B$, which are obtained by utilizing the projection operator technique.

| | First-order | Second-order |
|---|---|---|
| A | $k_y$ | $k_x^2,\ k_y^2, k_z^2, k_x k_z$ |
| B | $k_x, k_z$ | $k_x k_y, k_y k_z$ |

**TABLE S24.** The C-G coefficients between the *irreps* $\bar{E}^1$ and $\bar{E}^2$ through *irreps* $A$ and $B$.

| | $A\bar{E}^1$ | $A\bar{E}^2$ |
|---|---|---|
| $\bar{E}^1$ | 1 | 0 |
| $\bar{E}^2$ | 0 | 1 |

| | $A\bar{E}^1$ | $A\bar{E}^2$ |
|---|---|---|
| $\bar{E}^1$ | 1 | 0 |
| $\bar{E}^2$ | 0 | -1 |

| | $B\bar{E}^1$ | $B\bar{E}^2$ |
|---|---|---|
| $\bar{E}^1$ | 0 | 1 |
| $\bar{E}^2$ | 1 | 0 |

| | $B\bar{E}^1$ | $B\bar{E}^2$ |
|---|---|---|
| $\bar{E}^1$ | 0 | i |
| $\bar{E}^2$ | -i | 0 |

**SI References**